%% file: KIR.tex
\documentclass[12pt]{iopart}                        
\usepackage{tcolorbox}
\usepackage{amssymb}
\usepackage{bbm}
\usepackage{multirow}
\usepackage{ulem}
\usepackage[colorlinks=true,citecolor=blue,linkcolor=blue]{hyperref}
\usepackage{graphicx}
\usepackage{epsfig}

\newlength{\figwidth}
\newlength{\shift}
\newlength{\fight}
\newcommand{\fg}[3]
{
\begin{figure}[ht]
\vspace*{-0cm}
\[
\includegraphics[width=\fight]{#1}
\]
\vskip -0.2cm
\caption{\label{#2}
\small #3
}
\end{figure}}

\begin{document}
\title{Prospects and Applications Near Ferroelectric Quantum Phase Transitions:  A Key Issues Review}
\author{P. Chandra,$^{1}$  G.G. Lonzarich,$^2$ S.E. Rowley,$^{2,3}$ and 
J.F. Scott$^{2,4}$\footnote{The alphabetical ordering of the authors indicates they contributed equally to this article}}
\address{$^1$Center for Materials Theory, Department of 
Physics and Astronomy, Rutgers University, 
Piscataway, NJ 08854 USA.}
%\author{G. Lonzarich}
\address{$^2$Cavendish Laboratory, 
Cambridge University, J.J. Thomson Avenue, 
Cambridge CB3 0HE UK.}
%\author{S. Rowley}
%\author{J.F. Scott}
\address{$^3$Centro Brasileiro de Pesquisas Fisicas, Rua Dr. Xavier Sigaud 150, Rio de Janeiro 22290-180, Brazil.}
\address{$^4$Department of Physics, St. Andrews University, Scotland UK.}
\begin{abstract}
The emergence of complex and fascinating states of quantum matter in the 
neighborhood of zero temperature phase transitions suggests that such  
quantum 
phenomena should be studied in a variety of settings.
Advanced technologies of the future may be fabricated from materials
where the cooperative behavior of charge, spin and current can be manipulated
at cryogenic temperatures.  The progagating lattice dynamics of displacive ferroelectrics
make them appealing for the study of quantum critical phenomena that is characterized
by both space- and time-dependent quantities.  In this Key Issues article we 
aim to provide a self-contained overview of ferroelectrics near quantum phase transitions.
Unlike most magnetic cases, the ferroelectric quantum critical point can be tuned
experimentally to reside at, above or below its upper critical dimension; this feature
allows for detailed interplay between experiment and theory using both scaling and self-consistent
field models.  
Empirically the sensitivity of the ferroelectric $T_c$'s to external and to 
chemical pressure gives practical access to a broad range of temperature 
behavior over several hundreds of Kelvin.   
Additional degrees of freedom like charge and spin can be added 
and characterized
systematically.
Satellite memories, electrocaloric cooling and low-loss phased-array radar
are among possible applications of low-temperature ferroelectrics.  
We end with open questions for future research that include
textured polarization states and  
unusual forms of superconductivity that remain to be understood theoretically. 

\end{abstract}

%\pacs{1315, 9440T}
%
%\submitto{\RPP {\rm(invited)}}
%
%
%
\maketitle

\tableofcontents

\section{Introduction and FAQs}

\input Sec1_Intro.tex

\vspace{5 mm}

\section{Quantum Criticality Basics}

\input Sec2_QCB.tex

\section{Ferroelectrics Necessities}

\input Sec3_FEN.tex

\section{The Case of $SrTiO_3$ to Date}

\input Sec4_STO.tex

\section{A Flavor for Low Temperature Applications}

\input Sec5_LTAs.tex

\section{Open Questions for Future Research}

\input Sec6_OpenQs.tex

\section{Acknowledgements}

\input Sec7_Acks.tex

\pagebreak

\section{References}

%\bibliographystye{iopart-num}
\bibliographystyle{unsrt}
\bibliography{KIR}

\end{document}

%% file: Sec1_Intro.tex
%7/10/17
%
%Introduction

At first sight, the links between ferroelectrics, quantum 
phase transitions and quantum criticality may not be obvious.  After all, ferroelectrics 
are mostly non-metallic materials that are often studied towards specific functionalities 
at room temperature, whereas a key motivation for research in quantum
phase transitions and quantum criticality is their links with novel metallic behavior
and exotic superconductivity.  Our principal aim in this Key Issues article 
is to encourage more communication between researchers in these two mainly 
independent communities.  Let us begin by addressing frequently asked questions that 
might be posed by curious newcomers to the field 
in a colloquial fashion before presenting more detail in the subsequent 
parts of this article.

\vspace{5 mm}

\centerline{********************************}

\vspace{5 mm}

\noindent{\bf Aren't quantum fluctuations 
only important at $T= 0$ Kelvin ?} 

\vspace{3 mm}

\noindent Let's start by discussing 
what is meant by quantum fluctuations. 
We can begin by thinking about the amplitude 
fluctuations of a one-dimensional simple harmonic oscillator as a function 
of temperature, and let's take a look at Figure \ref{fig2} together. Here we see that, 
setting the constants $\hbar$ and $k_B$ to be unity, 
the important energy-scales are the 
temperature, $T$,  and the oscillator frequency $\Omega$.  
If $T$ is much greater than $\Omega$, then the variance in the amplitude, 
$\langle x^2 \rangle$, scales with $T$ and $\Omega$ drops out completely. 
In this case, the total variance results 
from purely classical (thermal) fluctuations and in Figure \ref{fig2}
their contribution to $\langle x^2 \rangle$ is indicated 
by a red line.  
However for lower temperatures, particularly in the interval 
$0 < T \lesssim \Omega$,  there is another contribution 
to $\langle x^2 \rangle$ above this classical red
line (see Figure \ref{fig2}) due to quantum fluctuations (blue line in Figure \ref{fig2}).
The total variance then becomes the sum of the quantum and the classical
components, where at $T=0$ only the quantum component survives.

\fight=5in
\fg{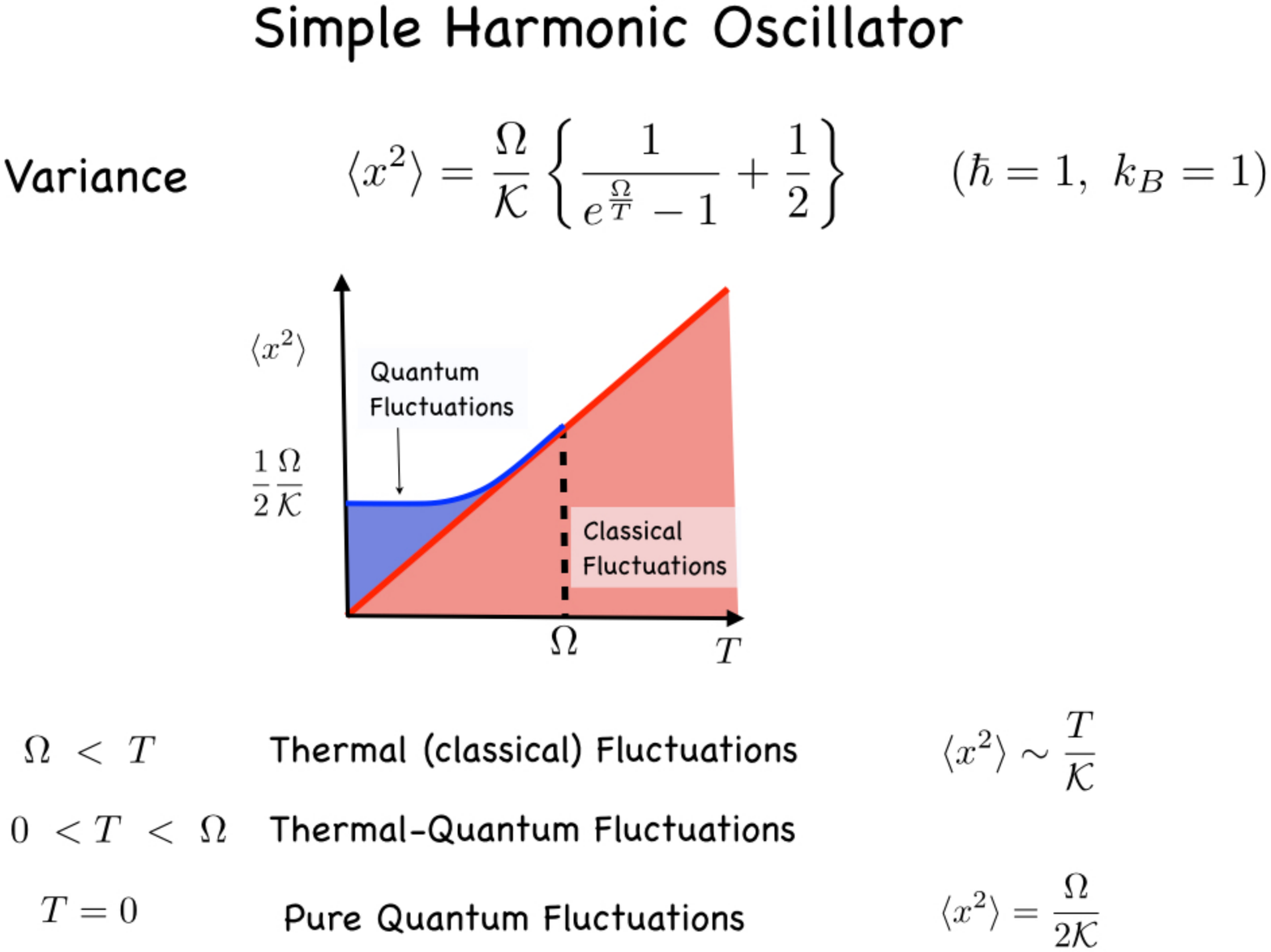}{fig2}{Amplitude variance of a simple harmonic oscillator where $\Omega$ and $\cal{K}$ are its frequency and stiffness respectively.}

\vspace{3 mm}

\noindent{\bf Fine, but what does this behavior of one simple 
harmonic oscillator have to do with quantum phase transitions?} 

\vspace{3 mm}

\noindent We are just getting to this conceptual connection.  Order 
parameter fluctuations play a key role at phase transitions, and we can consider the variance of each of 
their Fourier components one at a time. We can call each of these Fourier 
components a mode of wavevector $q$ whose 
behavior could be mapped onto that of single harmonic oscillator of amplitude $x$ where
$\Omega$ would be the oscillator frequency of the mode in question.  Now we are back to our Figure \ref{fig2} where
the full variance $\langle x^2 \rangle$ is plotted as a function of temperature for a particular mode of wavevector $q$.  
At a continuous phase transition the (mode) stiffness ${\cal K}$ vanishes for modes with wavevectors close to the ordering wavevector, so that the red and blue lines in Figure \ref{fig2}
becomes vertical and the amplitude fluctuations diverge.  If this occurs at a temperature $T \gg \Omega$, then the transition may be driven by
essentially classical fluctuations and is between high to low entropy states as a function of decreasing temperature. 
However at low temperatures where $0 < T \lesssim \Omega$, we have 
classical-plus-quantum (C+Q) fluctuations and here we are very interested in how these ``hybrid'' fluctuations lead to behavior and ordering distinct from those driven by 
their purely classical counterparts.  Again ${\cal K}$ at the ordering wavevector
goes to zero at the transition but now, in addition to the 
classical contribution, there is a quantum component to $\langle x^2 \rangle$.
Of course at strictly $T=0$ the fluctuations are purely quantum and the entropy change is zero for an
equilibrium system.  Therefore purely (equilibrium) quantum phase transitions are really transformations from one 
type of ordering to another. We emphasize this point because the term ``quantum disordered state'', that often appears in 
phase diagrams, is ambiguous and possibly confusing; it may only have useful meaning in cases where there is a finite
ground-state degeneracy in violation of the Third Law of thermodynamics.

\vspace{3 mm}  
  
\noindent{\bf Here you are telling us that quantum fluctuations increase amplitude fluctuations at low temperatures.  However Einstein and 
later Debye showed that quantum fluctuations reduced the specific heat from its classical value and that was a big success 
for the quantum theory.  How does this 
fit in with what you are saying?} 

\vspace{3 mm}

\fight=5in
\fg{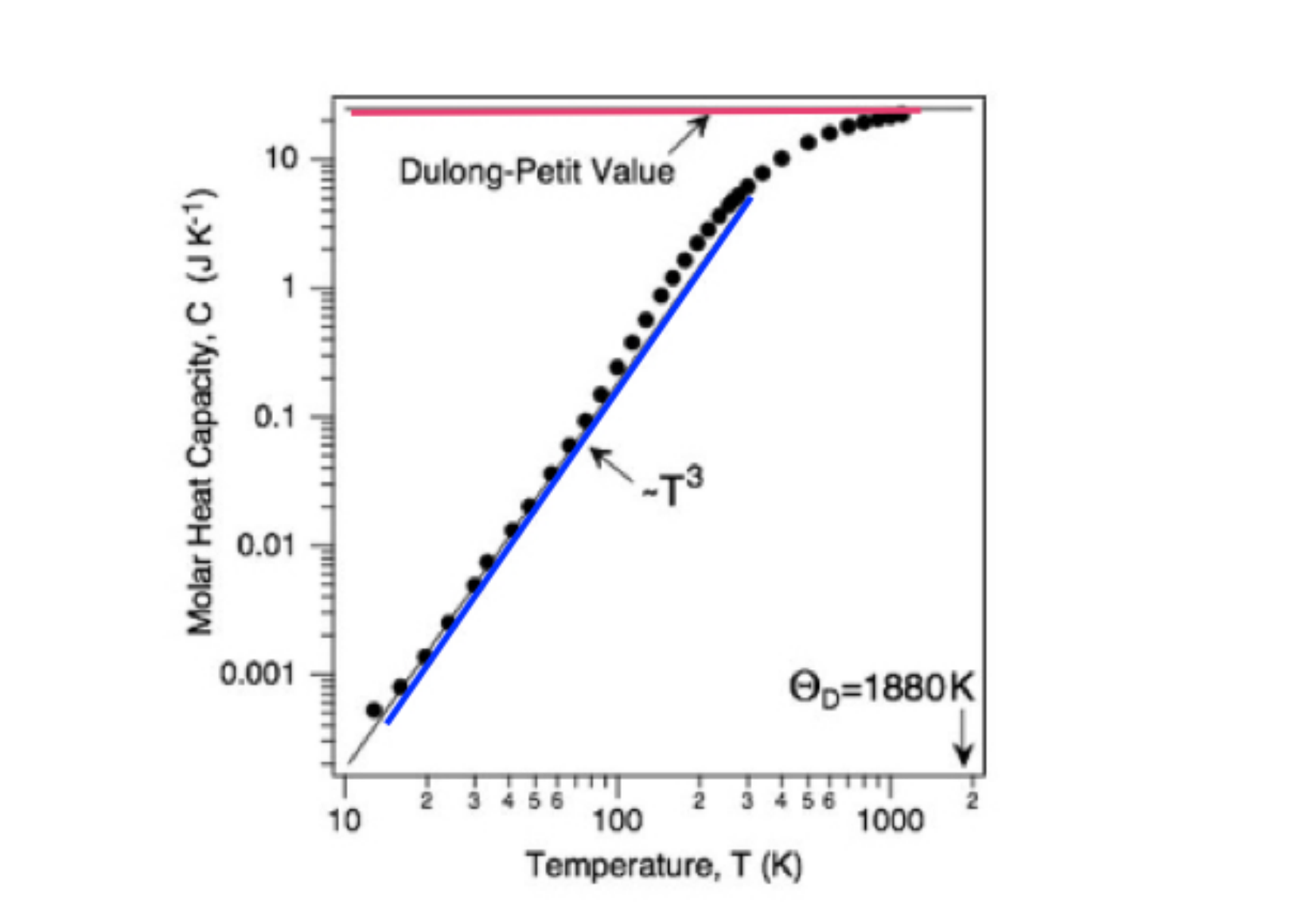}{fig3}
{Heat capacity of diamond vs. temperature. Note that 
at room temperature it is well below the classical Dulong-Petit value, indicating the importance of quantum effects at non-cryogenic temperatures. Adapted from 
Hofman \cite{Hofman15} with permission and thanks to P. Hofman.} 
%data points are from Desnoyer and Morrison \cite{Desnoyers58} and Victor \cite{Victor61}
\vspace{3 mm}  

\noindent You are of course completely correct that at low temperatures the specific heat of a solid is reduced compared to its 
classical constant value, and indeed this may seem counterintuitive given what we've just told you.  However we can in fact understand this behavior by looking 
again at Figure \ref{fig2}.  In our simple approach the energy is proportional to the variance of amplitude fluctuations, so the specific heat is then its derivative.  We see that 
the slope of the variance in the amplitude ($\langle x^2 \rangle$) is higher at temperatures $T >> \Omega$ than at $T << \Omega$, and indeed it is actually relatively flat in the 
approach to zero-temperature.  This means that the specific heat will be significantly lower at low temperatures compared to its constant value
at temperatures $T >> \Omega$ and we hope this answers the question. In Figure \ref{fig3} you see the specific heat $c_P$ of diamond that has a Debye 
temperature exceeding one thousand degrees (Kelvin); at room temperature $c_P$ is already temperature-dependent and thus the effects of quantum
fluctuations are observable without any fancy cryogenics!  

\vspace{2 mm}

\noindent As you suggest, the heat capacity is valuable in bringing out the dramatic quantum corrections to
classical behavior that can extend to room temperatures and above.  However it is also important to note that the heat capacity does not reflect the total
variance and depends only on the Bose function contribution; of course we are neglecting any temperature-dependence of $\Omega$ which would require a more
extended discussion. 

\vspace{3 mm}

\noindent{\bf So then why do we care about the total variance anyway if it isn't important for observable quantities?}

\vspace{3 mm}

\noindent We  agree that this is not obvious from our specific heat discussion. As we can see in Figure \ref{fig2},
the total variance has both classical and quantum components, where their relative contributions change as a function
of temperature.  Just as the classical part drives phase transitions for $T \gg \Omega$, it is the quantum part that drives
phase transitions for $T \ll \Omega$.  We should add that the total variance of the amplitude fluctuations can be probed, for example, 
by neutron scattering experiments where the neutron loses energy to the system so that both the zero-point and the 
Bose function contributions are measured.  Again we stress that it is the total variance that is crucial for the ``disruption''
of the initial form of order.

\vspace{3 mm}

\noindent{\bf What does quantum criticality mean?}
 
\vspace{3 mm}

\noindent In a nutshell, quantum criticality refers to a second-order phase transition that occurs at zero temperature.  More generally, it's 
probably easiest to answer your question by comparing quantum criticality
to its classical counterpart.  At a continuous phase transition the inverse order parameter susceptibility vanishes so that
the order parameter correlation function becomes scale-invariant.  This means that it decays with distance and time not
exponentially but rather gradually in a power-law form. The thermodynamic variables depend only on  
scale-invariant correlation functions in space for classical criticality, but crucially on both space and time for quantum
criticality.  This leads to new critical exponents that are quantum in nature depending on details of the order parameter 
dynamics. 

\vspace{3 mm}

\noindent{\bf In ferroelectrics classical criticality is difficult to observe in practice. Why isn't the same true for quantum criticality?}
 
\vspace{3 mm}

\noindent As you suggest, the criteria for observing classical and quantum criticality are quite different.
For example classical criticality just below $T_c$ is defined as 
the region near a finite temperature phase transition where 
fluctuations in the order parameter are comparable to the 
average of the order parameter itself.  Empirically it has been found 
that mean-field theory works very well near classical ferroelectric
phase transitions, though of course most are first-order.
Actually many ferroelectrics reside close to tricritical points at ambient pressure.  Therefore it's not surprising that pressure-tuned ferroelectric 
transitions are continuous, at least in practice. More generally, continuous ferroelectric quantum phase transitions are expected if one is willing to 
tune not only pressure or composition but also the electric field. As an aside, we should also note that textured states are known to reside near first-order quantum phase transitions, so they can be very interesting too.

\vspace{3 mm}

\noindent{\bf What defines 
the quantum critical region?} 

\vspace{3 mm}

\noindent It's important to realize that temperature is {\sl not} a simple 
tuning parameter 
at a quantum phase transition.  Indeed temperature provides the low-energy 
cutoff for quantum fluctuations where the associated time-scale is defined through the Heisenberg uncertainty relation $\Delta t \sim \frac{\hbar}{k_B T}$.
In this sense temperature plays the role of a finite-size effect in time at a quantum critical point. The quantum critical region is defined by the interplay 
between the scale-invariant order parameter fluctuations and the temporal boundary conditions imposed by finite temperature; most importantly it is
accessible experimentally with distinct observable signatures.

\vspace{3 mm}  

\noindent{\bf Now can you please explain  why $d+1$ is the effective dimension?}

\vspace{3 mm}

\noindent In the
case of purely classical fluctuations, the amplitude for each mode of wavevector $q$ depends only
on the temperature and not on its dynamical properties, as we've already noted.  Therefore
its statistical mechanical description involves only the $d$ dimensions of wavevector (or of
real) space.  However when quantum fluctuations are present, the mode frequency as well as the 
temperature are important for the statistical mechanical characterization; for example see the
expression for the variance in Figure \ref{fig2}.
In general there is a distribution of frequencies $\omega$ associated with each mode that reduces
to a $\delta$-function in the special case of a simple harmonic oscillator where $\omega = \Omega$.  
More generally each mode has a power spectrum distribution of frequencies that results in 
a statistical mechanical description involving not only the sum over wavevectors but also over
frequency $\omega$.  The effective number of dimensions to be associated with
the dynamics is dependent on the frequency-wavevector dispersion relation.
If the dispersion is linear, as it is for ferroelectrics, space and time
enter the statistical description on equal footing leading to an overall effective dimensionality
of ``d + 1'' referring to $d$ space and $1$ time dimensions.  Another subtlety is that the effective time 
dimension is of finite size except in the limit
$T \rightarrow 0$ as we've just discussed.

\vspace{3 mm}

\noindent{\bf New functionalities are of great interest to the ferroelectrics 
community, so are there useful low-temperature applications for these 
materials that could be pursued in parallel to studies of quantum criticality?}

\vspace{3 mm}

\noindent The trends for future devices are faster, 
lighter and smaller. Ferroelectric films are used as both active 
and passive memory elements where data is stored as the presence 
(or absence) of charge.  
Reduced operating temperatures lead to lower leakage currents and to
increased breakdown fields, both crucial for keeping competitive
with faster access and high-density needs.   

\vspace{2 mm}

\noindent Electrocaloric cooling, the change in temperature with applied 
electric field, could be developed to access cryogenic temperatures 
just as its magnetic counterpart, magnetocaloric cooling, is
often used to access millikelvin temperatures and below.

\vspace{3 mm}

\noindent{\bf There was some work exploring cryogenic
electrocaloric cooling some time ago that was not pursued as the observed effects were too small for practical use...what has changed since then to make you optimistic about this application?}

\vspace{3 mm}

\noindent In a nutshell, current thin-film and multicapacitor 
technologies means that we can increase breakdown fields, particularly at 
low temperatures without loss of effective volume. It is certainly much 
easier and cheaper to apply electric rather than magnetic fields, and 
we'll have more to say about electrocaloric cooling shortly.

\vspace{2 mm}

\noindent We should also note that the radiation-hardness of ferroelectric 
memories makes them ideal for satellite applications where there is 
repeated passage through the Van Allen belts and naturally cold
temperatures! Indeed in efforts to develop radiation-tolerant electronics,
NASA has performed on-orbit tests of ferroelectric random access
memories, FRAMs, on micro-satellites (see Figure \ref{fig4}). Furthermore 
NPSAT1, a small satellite built by the Naval Postgraduate School with 
FRAMs on board, is due to launch on the SpaceX Falcon Heavy sometime in 2017.

\fight=5in
\fg{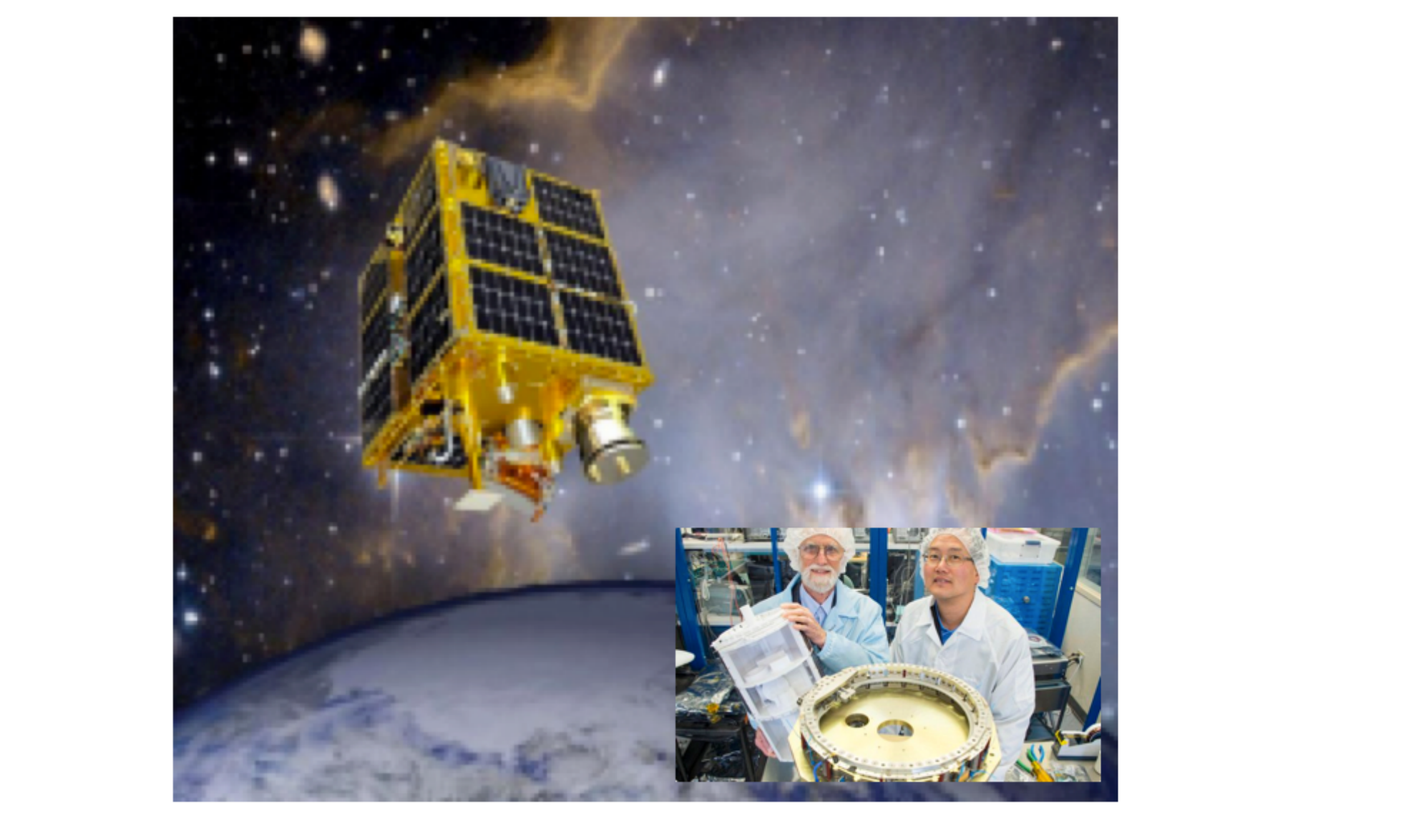}{fig4}{(Left) Artist's rendition of NASA's 
Fast and Affordable Science and Technology Satellite
(FASTSAT)  
with ferroelectric randon access memory (FRAM) for radiation robustness
reprinted from MacLeod et al. \cite{MacLeod11,MacLeod12} with permission and thanks to T.C. MacLeod; (Inset) 
Naval Postgraduate School scientists R. Panholzer and D. Sakoda with 
several structural pieces of Naval Postgraduate School Satellite 1 with FRAM 
\cite{nps} due to launch on a STP-2 mission in 2017 
 on a SpaceX Falcon Heavy rocket \cite{falcon} (US Navy Photo by Javier 
Chagoya, reprinted from \cite{nps} 
with permission and thanks to J. Chagoya and the NPS Public Affairs Office).}

\vspace{2 mm}

\noindent Another potential application for cryogenic ferroelectrics is 
in phased array radar that would replace large, heavy radar antennae 
that mechanically rotate. Beam steering would be achieved electrically
by varying the phase of a voltage train with a field-tuned LC circuit.
In order for such array radar devices to be competitive with their
mechanical analogues the dielectric loss must be very low, about $0.01 \%$,
and thus this should be a niche for cryogenic ferroelectrics.  We should
point out that the entire device would not have to be at low 
temperatures...on-chip electrocaloric cooling for the capacitor could do 
the job nicely!

\vspace{3 mm}

\noindent{\bf So it sounds like there are several low-temperature applications for ferroelectrics that can be explored.  Now back to a more general question. What from our knowledge of magnetism can be 
transferred to ferroelectricity?}

\vspace{3 mm}

\noindent There are indeed similarities between ferroelectrics and
ferromagnets, but there are also key differences.  For example, the 
polarization is a classical object and thus is not quantized in contrast to
the spin in a magnet.  
%Jim's input
%Furthermore the polarization is not localized, rather it extends over a 
%whole unit cell and is a spatial average over all ions there.  
%For example, the unit cell changes with temperature, due to thermal expansion, 
%and this affects the temperature-dependence of the polarization.
%
Crystal fields lead to strong anisotropy in
ferroelectrics whereas magnetic anisotropy is usually orders of
magnitude smaller and is principally due to spin-orbit coupling; this leads
to different domain structures in these two distinct classes of materials.
The dynamics in ferroelectrics are dominated by propagating vibrational modes, 
whereas in magnets there is spin precession.  These are just some of the 
reasons one has to be careful going back and forth between magnetism and 
ferroelectricity, and we'll be discussing this in more detail shortly.

\vspace{3 mm}

\noindent{\bf Most of our experiments in quantum criticality are on 
metallic systems and most ferroelectrics are insulating.  So where is the 
common ground?}

\vspace{3 mm}

\noindent We usually emphasize the fact that ferroelectrics are analogous to ferromagnetic
insulators.  However in the present context, they have interesting features in common with 
itinerant magnets.  In a ferroelectric at high temperatures, the polarization is not well-defined due to dynamical
fluctuations in the separation between charges.  Similarly in an itinerant magnet, the magnetic
moment is not well-defined at high temperatures since the number of electrons in a unit cell is
constantly fluctuating.  So in that sense the two are not that different.  We should add that there also
have been studies of doped bulk strontium titanate that indicate very interesting metallic and
superconducting behaviors.  Indeed doped strontium titanate is the superconductor with one of
the lowest carrier densities known to date.  Its Fermi temperature is lower than its Debye temperature, a feature also 
seen in many heavy fermion superconductors.  Thus it most probably cannot be described by a conventional
theory of superconductivity.  

\vspace{3 mm}

\noindent{\bf So, given our discussion, what can ferroelectricity bring to the
study of quantum criticality?}

\vspace{3 mm}

\noindent Empiricially the sensitivities of the ferroelectric transition temperatures to pressure are remarkable!  As an example, in order to cover 300 K changes in magnetic $T_c$'s, we must usually apply hundreds of kilobars, whereas in ferroelectrics the same temperature range can be achieved with more than a factor of ten less in pressure. Furthermore the electric field as another tuning 
parameter offers tremendous
advantages over its magnetic counterpart, as an electric field is 
significantly easier to apply and doesn't require
a lot of extra coils, special cells etc.  Also, through gated control of carriers, there is another type
of continuous fine-tuning available without the need for multiple samples at different doping levels. 
In the quantum regime, as we discussed earlier, a system's thermodynamic behavior involves both space and 
time and hence dynamics; since the dynamics of ferroelectrics and ferromagnets are different, their
quantum critical behavior will also be distinct. More generally, another 
class of materials for experiment is crucial as we collectively explore the possibility of universality in quantum critical 
phenomena.

\vspace{5 mm}

\centerline{********************************}

\vspace{5 mm}

So we see, there is quite a lot to discuss!  We note that there has been tremendous ``historical
entanglement'' here between the fields of ferroelectrics and criticality; the 
first logarithmic corrections to mean-field exponents due to fluctuations at marginal dimensionality were calculated for a uniaxial ferroelectric \cite{Larkin69}.  Similarly the transverse-field Ising model, 
one of the simplest models demonstrating a quantum phase transition, was first developed to describe a transition 
transition in the ferroelectric potassium dihydrogen phosphate $KH_2PO_4$ (often denoted as KDP) \cite{Chakrabarti96}.  
Indeed historically there have been several ``waves'' of interest in low-temperature
paraelectrics that are not completely chronologically distinct; here, in the interest of compactness,  we refer the 
interested reader to previous reviews to discuss 
these developments \cite{Samara01,Kvyatkovskii01}.  
In the 1950s, perovskites like $SrTiO_3$
and $KTaO_3$ were of experimental interest since their dielectric properties were so different
from those of (ferroelectric) $BaTiO_3$.  Next, in the late 60s through the mid-80s,
with the development of renormalization group, they were settings
to test lattice model calculations of quantum critical exponents and to study the importance
of long-range dipolar interactions in different dimensions.  More recently there has been tremendous
interest in the interplay of polarization with other degrees of freedom, so there has
been much effort towards modelling phase diagrams of materials for a wide range of temperatures
with the aim of raising interesting low-temperature phases to room temperature for appropriate
applications \cite{Spaldin10}. A closely related field is that of ferroelastics, the mechanical analogue of ferroelectricity and ferromagnetism, that is associated with shape memory effects \cite{Salje12}.

\vspace{2 mm}

In this article, we'd like to encourage yet another ``wave'' of interest in the low temperature
behavior of paraelectrics/ferroelectrics, one motivated by the quest to discover new quantum
states of matter near quantum phase transitions \cite{Sachdev99,Continentino01,Gegenwart08,Sachdev08}. 
Materials near their displacive ferroelectric quantum transitions are particularly elegant examples
of quantum criticality \cite{Rechester71,Khmelnitskii73,Roussev03,Palova09,Das09,Riseborough10,Rowley14} with few degrees of freedom
and propagating dynamics that distinguish them from their magnetic counterparts.  Furthermore, as we'll discuss, 
they are dimensionally tunable so they can be studied experimentally and theoretically at, above and below
their upper critical dimensions. Additional degrees of freedom like spin and 
charge can be added and characterized systematically in these materials, leading
to rich phase behavior as yet mostly unexplored.

\vspace{2 mm}

Let's not get ahead of ourselves.  To ensure
that everyone is roughly on the same page, we aim for a self-contained article with
many references.  We apologize in advance to any researchers whose work has been inadvertently 
overlooked, and  we hope that our bibliography will give the interested reader a good starting point to 
explore topics of interest  in more depth. We begin with ``Quantum Criticality Basics'' in Section II and then
continue in III to ``Ferroelectrics Necessities.''  
Then (IV) we  discuss the specific case of the material $SrTiO_3$ and its
behavior at low temperatures.  ``A Flavor for
Low Temperature Applications'' is the next section (V) and we end (VI) 
with several open questions for future research.

%% file: Sec2_QCB.tex
%7/10/17
%
%Quantum Criticality Basics

Our aim here is to present key ideas of quantum criticality with minimal 
formalism to those new to the field, using familiar concepts whenever 
possible;  naturally we refer the reader eager for more detail to a 
number of excellent reviews \cite{Sachdev99,Continentino01,Gegenwart08,Sachdev08,Vojta02,Vojta03,Coleman05} 
In particular our focus will be the temperature behavior
of observable quantities near a quantum critical point, eventually associated
with ferroelectricity; this goal will guide our discussion.
We are all familiar with classical phase transitions 
where the order parameter develops at a characteristic critical temperature.  
This  standard picture assumes purely classical (thermal) 
fluctuations which is certainly appropriate for the temperatures of general
interest. 
As we've just discussed in the Introduction, 
quantum fluctuations also contribute to order parameter fluctuations of
modes with characteristic frequencies of the order of or greater than
the temperature; here for presentational simplicity we have set the constants 
$\hbar = k_B = 1$.  
However if, as $ T \rightarrow 0$, 
the fluctuation-selection of different ground states is 
enhanced by another tuning parameter, $g$, then there 
is the possibility of a $T=0$ continous quantum phase transition.

Let's resume our previous discussion of order parameter fluctuations
where we treated each Fourier mode as 
a simple harmonic oscillator of amplitude $x$ with frequency
$\Omega$. The total variance in the mode amplitude is then
\begin{equation}
\langle x^2\rangle = \left\{n_{\Omega} + \frac{1}{2} \right\} \Omega \ \chi
\label{shovar}
\end{equation}
where $n_{\Omega}$ refers to the Bose function and 
$\chi = \frac{1}{\cal K} \ (= {\rm Re} \ \chi_{\omega=0})$ where 
${\cal K}$ is the relevant spring stiffness or elastic constant. 
We recall that for a simple harmonic oscillator
\begin{equation}
\label{imsho}
{\rm Im} \ \chi_{\omega} = \frac{\pi}{2} \ \omega \ \chi \ \delta(\omega - \Omega) \quad\quad (\omega > 0) 
\end{equation}
so that 
we can rewrite (\ref{shovar}) as 
\begin{equation}
\langle x^2 \rangle = \frac{2}{\pi} \int_0^{\infty} d\omega \left \{n_\omega + \frac{1}{2} \right\} {\rm Im} \ \chi_{\omega}.
\label{shovar2}
\end{equation}
We note that this link between the variance of amplitude fluctuations and the 
imaginary part of the response, here derived
for a simple harmonic oscillator, is actually a much more general result 
associated with the 
fluctuation-dissipation (Nyquist)
theorem \cite{Landau80}.

We can generalize (\ref{shovar2}) to a sum over all modes labelled by 
wavevector $q$, for example, in the entire Brillouin zone.
Let us now transition to the amplitude of the scalar order parameter $\phi$ 
that here is a (dipole) moment density that can be either magnetic or 
electric; we use this terminology for simplicity
to avoid confusion with other common symbols often associated with 
pressure.  Then, following our 
previous argument, the variance of the amplitude fluctuations
of the moment is  
\begin{equation}
\langle \delta \phi^2 \rangle 
= \frac{2}{\pi} \sum_q \int_{0}^{\infty} d \omega 
\left\{n_\omega + \frac{1}{2}\right\} \quad {\rm Im} \ \chi_{q\omega}
\label{ggfdt}
\end{equation}
where $\phi = \overline{\phi} + \delta \phi$, $\overline{\phi}$ is the average, $\langle \delta \phi \rangle = 0$ and 
\begin{equation}
{\rm Im} \ \chi_{q\omega} = \frac{\pi}{2} \ \omega \ \chi_q \ \delta (\omega - \omega_q)  \quad \quad (\omega > 0)
\label{imchi}
\end{equation}
in the propagating limit 
where $\omega_q$ is the oscillator frequency of the mode of 
wavevector $q$; naturally more general power spectra are 
also possible \cite{Coleman15}.

Equation (\ref{ggfdt}) is composed of a strongly temperature-dependent contribution $\langle \delta \phi_T^2 \rangle$ involving the 
Bose factor $n_\omega$; the remainder ($\langle \delta \phi_{ZP}^2 \rangle$ involving the factor $\frac{1}{2}$ instead of $n_\omega$) is 
due to ``zero-point'' fluctuations.  
Here we focus on $\langle \delta \phi_T^2 \rangle$ since it is dominant in 
determining 
the temperature-dependence of the observable properties of interest here.  We note that the zero-point contribution 
mainly affects the $T=0$ properties and as noted previously can drive a 
quantum phase transition; in particular here it is assumed just to 
renormalize the underlying 
parameters of the free energy-energy expansion in the vicinity of
the zero-temperature transition \cite{Landau80} 
that we'll present shortly. 
Let us now return to equation (\ref{ggfdt}). 
At high temperatures ($T >> \omega$), $n_\omega \approx \frac{T}{\omega}$; invoking causality in the 
form of the Kramers-Kronig relations, 
we obtain a generalized equipartition theorem \cite{Landau80} 
\begin{equation}
\label{cfdt}
\langle \delta \phi_T^2 \rangle \approx \ T \sum_{q < q_{BZ}} \ \chi_q \quad 
\quad \quad (T >> \omega_q \quad {\rm for}  \ q < q_{BZ}).
\end{equation}
Here we see that the dynamics drop out completely 
of the classical equilibrium description.
We also note that in (\ref{cfdt}) we have a d-dimensional wavevector summation over the Brillouin zone
that implies a d-dimensional theory in real space.

\fight=4.5in
\fg{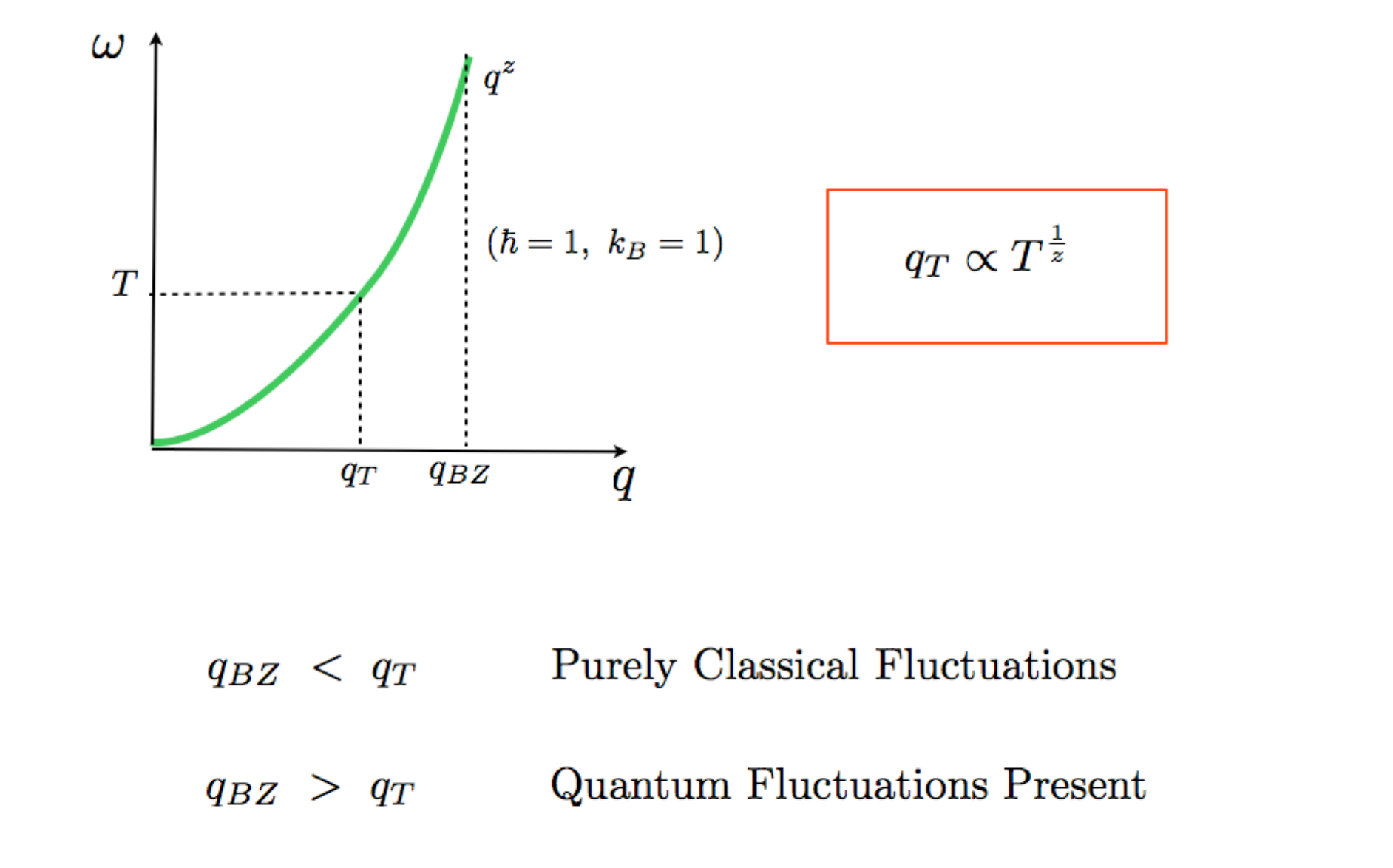}{fig5}{Important wavevectors and the dispersion $\omega \propto q^z$.}

By contrast, in the regime ($T << \omega$) 
where quantum effects are important, 
$n_{\omega} \approx e^{-\frac{\omega}{T}}$ and the dynamics remain.
In order to proceed with our treatment of (\ref{ggfdt}), 
we therefore must consider the dispersion $\omega_q$;
please see Figure \ref{fig5}. In particular we'll get a purely classical result,
(\ref{cfdt}), if
all the modes in the Brillouin zone are excited; otherwise the modes will be
classical up to a wavevector cutoff determined by quantum 
mechanics (see Figure \ref{fig5}).
The relevant wavevector scales are the Brillouin zone ($q_{BZ}$) and the thermal ($q_T$) wavevectors,
where the latter's temperature-dependence,  
via the dispersion $\omega_q \propto q^z$ for low $q$, is
\begin{equation}
q_T \propto T^{\frac{1}{z}}
\label{qt}
\end{equation}   
and we note that $\frac{1}{q_T}$ is a generalized deBroglie wavelength that
correponds to the usual free-particle case when $z=2$.
We emphasize that the smaller of the two wavevector scales $q_T$ and $q_{BZ}$ 
serves as a cutoff for the classical fluctuations.
If $q_T < q_{BZ}$ then not all modes in the Brillouin zone are 
thermally excited; in this case the dynamical exponent enters (\ref{ggfdt}) via $q_T$ and
thus quantum effects contribute to the variance of the order parameter fluctuations.

Let us now apply these ideas towards analyzing (\ref{ggfdt}) when the important cutoff is $q_T$. 
We revisit the most strongly temperature-dependant part of the
$\omega$-integral in (\ref{ggfdt}), breaking 
it up into two separate parts as approximately
\begin{equation}
\label{qfdt1}
{\cal I}  \ = \ {\cal I}_1 \ + \ {\cal I}_2 \ \approx \  \int_0^T d\omega \left(\frac{T}{\omega}\right) \ {\rm Im} \ \chi_{q\omega} + \int_T^\infty \ d\omega \ e^{-\frac{\omega}{T}} \ {\rm Im} \ \chi_{q\omega}.
\end{equation}
We note that for $q < q_T$ the delta function in (\ref{ggfdt}) and (\ref{imchi}) ensures that  
only ${\cal I}_1$ contributes in (\ref{qfdt1}); for $q > q_T$, ${\cal I}_1$ is zero and ${\cal I}_2$ involves an 
exponential damping factor and thus can be ignored to leading order.
Therefore, using Kramer-Kronig relations,   
we can write (\ref{ggfdt}) as
\begin{equation}
\label{qfdt}
\langle \delta \phi_T^2 \rangle \ 
%\approx 
%\frac{2}{\pi} \sum_{q < q_T}  \ 
%\int_0^T \left(\frac{T}{\omega}\right) \ {\rm Im} \ \chi_{q\omega} 
\approx
\ T \ \sum_{q < q_T} \ \chi_q    \quad \quad \quad  (T << \omega_q \quad {\rm for}  \ q < q_T).
\end{equation}
where the dynamics are present via  (\ref{qt}).  
In this approach, the key distinction
between the two moment variances, (\ref{cfdt}) and (\ref{qfdt}), 
lies in their wavevector cutoffs:  
in the purely classical case (\ref{cfdt}) it is a constant ($q_{BZ}$), 
whereas when  quantum effects are important, (\ref{qfdt}), 
the dynamical exponent $z$ enters through $q_T$. 

Using the Landau theory of phase transitions (also called the 
Landau-Devonshire theory in the area of ferroelectric phase 
transitions) \cite{Landau80,Lines77,Chandra07} 
combined with (\ref{cfdt}) and (\ref{qfdt}), we can
relate the variance $\langle \delta \phi_T^2 \rangle$ to the susceptibility $\chi$, an observable quantity \cite{Rowley14,Morice16}.
In the magnetic and ferroelectric cases of interest here,
\begin{equation}
\label{chiform}
\chi_q^{-1} \propto {\kappa^2 + q^2}
\end{equation}
where $\kappa$ is the inverse correlation length
so that in the limit of $q \rightarrow 0$ we have
\begin{equation}
\chi^{-1} \propto {\kappa^2}.
\label{chi0}
\end{equation}
We recall that Landau theory is a symmetry-based description 
of macroscopic properties near a phase transition; here we will be considering
behavior on length-scales greater or equal to $\frac{1}{q_T}$.
This coarse-graining ensures that the main effects of 
zero-point fluctuations are absorbed in the Landau coefficients but that 
thermal effects show up through the fluctuations of the order 
parameter field coarse-grained over $\frac{1}{q_T}$. We assume 
that this scale is large enough so that a Taylor 
expansion of the free energy is still reasonable for our applications.

The Landau free energy density for a system with moment $\phi$ and conjugate field 
$\cal{E}$ is
\begin{equation}
f  = \frac{1}{2} \alpha \phi^2 + \frac{1}{4} \beta \phi^4 + 
\frac{1}{2} \gamma |\nabla \phi|^2 - {\cal E} \phi 
\label{fLandau}
\end{equation}
where $\alpha \rightarrow 0$ at the transition and 
$\beta$ and $\gamma$ are positive constants for a continuous phase transition to a uniformly ordered state that we wish to consider.
Minimizing this free energy with respect to the order parameter
$\phi$, we obtain
\begin{equation}
{\cal E} = \alpha \phi + \beta  \phi^3 - \gamma \nabla^2 \phi.
\label{emf}
\end{equation}

Solving for $\phi$ in (\ref{emf}), we obtain its most probable 
value associated with the maximum 
of its probability distribution.  
In order to determine the observed moment,
we consider the effects of fluctuations due to a random (Langevin) field
added to ${\cal E}$. More specifically we must  
average over the random fluctuations in (\ref{emf})
using 
$\phi \rightarrow \overline{\phi} + \delta \phi$ 
where $\overline{\phi}$ is the average and $\langle \delta \phi \rangle = 0$;
we obtain 
\begin{equation}
{\cal E} = (\alpha + 3 \beta \langle \delta \phi^2 \rangle) \overline{\phi}  +  
\gamma \nabla^2 \overline {\phi}
\end{equation}
to lowest order where we note that the variance term arises from the 
anharmonic effects
of the cubic term in the equation of state. 
In the limit of small $\overline{\phi}$ and ${\cal E}$, we can Fourier transform 
this expression to obtain
\begin{equation} 
\chi_q^{-1}  = (\alpha + 3 \beta \langle \delta \phi^2 \rangle)   + q^2.
\label{Efluct0}
\end{equation}
Taking the expression  (\ref{Efluct0})
in the $q \rightarrow 0$ limit and again retaining the most strongly
temperature-dependent terms,
we find that 
\begin{equation}
\lim_{T\rightarrow 0} \kappa^2 \propto \langle \delta \phi_T^2 \rangle
\label{chi}
\end{equation}
where we have assumed a quantum critical point (QCP) so that
$\alpha \rightarrow 0$ as $T \rightarrow 0$.

The careful reader may ask why we are distinguishing between the most probable 
and the average (observed) value of $\phi$, and this question can be addressed by discussion of equation (\ref{Efluct0}).  If the coarse-graining underlying our Landau theory is macroscopic, then the $q$ phase space and thus the variance is small, except in the Ginzburg regime to be defined below, so that the the most probable and the average values are essentially identical.  However, as we have already noted, our coarse-graining is 
mesoscopic and not macroscopic and therefore we must include the variance in 
our calculations.  An alternative way to address this issue is to recall that 
the true equation of state is found by averaging over the most 
probable one (\cite{Feynman65}); for
a Gaussian theory of course the average and the most probable values of $\phi$
are identical. 
Finally we emphasize that (\ref{chi}) is only valid near a $T_c = 0$ phase transition since for a nonzero $T_c$
there are additional terms proportional to $T_c \neq 0$
so that this expression of proportionality no longer holds \cite{Morice16}.

We can now combine (\ref{qfdt}), (\ref{chiform}) and 
(\ref{chi}) to determine the 
temperature-dependence of the susceptibility near a quantum critical point; 
towards this goal, we write
\begin{equation}
\kappa^2 \propto \sum_{q<q_T} \frac{T}{\kappa^2 + q^2} 
\approx T \int_{\kappa}^{q_T} \frac{q^{d-1}}{q^2} \approx
T\ q_T^{d-2}  \ \  \left\{1  - \left(\frac{\kappa}{q_T}\right)^{d-2}\right\}.
\label{kappa1}
\end{equation}
where, using $q_T \propto T^{\frac{1}{z}}$,  we are tempted to 
neglect the $\frac{\kappa}{q_T}$ term on the right-hand side of (\ref{kappa1})
and write
\begin{equation}
\chi^{-1} \propto \kappa^2 \propto T^{\frac{(d + z - 2)}{z}}.
\label{chiqt}
\end{equation}  
(\ref{chiqt}) shows that the quantum critical exponent for the susceptibility is
$\frac{d+z-2}{z}$ that can be compared to the classical value of unity 
(e.g. the Curie susceptibility)
outside the Ginzburg regime.
Now we can ask, when is this approach valid?  We can answer this question
by rearranging (\ref{kappa1}) to yield
\begin{equation}
\left(\frac{\kappa}{q_T}\right)^2 \propto T^{\frac{(d+z-4)}{z}} 
\left\{ 1 - \left(\frac{\kappa}{q_T}\right)^{d-2} \right\}. 
\label{kappa2}
\end{equation}
From (\ref{kappa2}), we see that $\left(\frac{\kappa}{q_T}\right) 
\rightarrow 0$ as
$T \rightarrow 0$ if $d_{eff} \equiv d + z > 4$; 
in this case the inverse susceptibility in the approach to a QCP has the 
temperature-dependence displayed in (\ref{chiqt}) and no further fluctuation
effects need to be considered. $d^{upper}_{space} = 4 - z$ is thus the upper critical spatial dimension of this theory.  An analogous treatment leads to 
$d^{upper} = 4$ for the purely classical description 
\cite{Continentino01,Morice16}; it is more 
complicated than the $T \rightarrow 0$ case due to the presence of 
more finite terms, so here we will simply state the result.

Let us now return to (\ref{kappa1}) and (\ref{kappa2}) with cutoff $q_T$.  It is as if the frequency (or time) dimension is equivalent
to $z$ wavevector (or space) dimensions through the dispersion relation
that relates frequency to $z$ factors of wavevector ($\omega \propto q^z$).  
Perhaps it is easier 
to state that the inclusion of
dynamics in quantum critical phenemona theory reduces the upper critical 
dimension from 4 in the classical
limit (where dynamics can be ignored) to $4-z$ (where dynamics must be 
considered).  From this standpoint,
we are usually above the upper critical dimension at a quantum phase 
transition whereas we are below it
for its classical counterpart.

We have already noted that the frequency dimension is truncated by the Bose 
function and can be envisioned to have a finite-size of order $T$, so that the corresponding time dimension
is of finite-size of order $\frac{1}{T}$.  The crucial point here is that the role of temperature
near a quantum critical point is to constrain the temporal dimension; for $d < d^{upper}_{space} = 4 - z$, 
thermal effects can be treated compactly via the ideas of finite-size scaling.  
More generally, we note that the frequency integration in (\ref{ggfdt}) 
can be performed by contour integration where the poles for the Bose function are imaginary \cite{Coleman15}. This is an alternative 
to the real-frequency and real-time description given here, and it yields the same results mathematically.

\vskip0.2in

\input Sec2_Box_QC_Highlights

\vskip0.2in

%% file: Sec2_Box_QC_Highlights
\begin{tcolorbox}

\center{\bf Quantum Criticality:  Key Concepts}
\vskip0.20in
\begin{itemize}
\vskip0.10in

\item The {\bf dynamical} properties of the order-parameter fluctuations affect
the equilibrium thermodynamic properties in the quantum critical regime (in contrast to their classical counterparts where only thermodynamic properties usually only depend on statics).

\item {\bf The dynamical exponent $z$}, defined by the dispersion relation 
($\omega \propto q^z$) at the quantum critical point, plays an important role in quantum critical phenomena.

\item The {\bf effective dimensionality}, $d_{eff} = d + z$, is the sum of the spatial and temporal dimensions, where the latter is represented by the dynamical exponent.

\item Near a quantum critical point (QCP), {\bf temperature acts as a boundary condition on time} and {\bf not} as a simple tuning parameter.

\item There exists a {\bf finite-temperature quantum critical region} near a QCP where there is a gapless dispersion, $q_T << q_{BZ}$ and $q_T \propto T^{\frac{1}{z}}$.

\item At sufficiently low temperatures near a QCP, the {\bf temperature-dependence of the inverse susceptibility} is
\begin{equation*}
\nonumber
\chi^{-1} \ \propto \ T^{\frac{d + z -2}{z}}  \quad \quad (d + z > 4)
\end{equation*}
(with weak logarithmic corrections for $d + z = 4$)
\end{itemize}

\end{tcolorbox}

%% file: Sec3_FEN.tex
%8/7/17
%
%Ferroelectrics Necessities

So why study the influence of quantum effects in materials
with ferroelectric tendencies?  Before addressing this question,
let us familiarize ourselves with key features of ferroelectrics (FE);
here we emphasize aspects important to our topic at hand, referring
the reader eager for further information to several detailed
reviews and books \cite{Kvyatkovskii01,Lines77,Fatuzzo67,Jona93,Kittel96,Strukov98,Scott00,Rabe07,Tomic15,Benedek16}. 

From a working ``engineering'' standpoint, a ferroelectric is a 
material that has a spontaneous polarization that is switchable by 
an electric field of practical magnitude; in a finite
system the polarization is defined as the dipole moment per volume averaged
over the unit cell volume \cite{Rabe07}. In Figure \ref{fig6}, the link between 
ferroelectrics, pyroelectrics, piezoelectrics and dielectrics is presented 
graphically. In piezoelectrics an applied mechanical stress 
results in a voltage and vice versa \cite{Lines77,Fatuzzo67,Jona93,Scott00}.  
A change in temperature causes an electrical polarization in
a pyroelectric \cite{Lines77,Fatuzzo67,Jona93,Scott00} and it is the 
practical switchability of this polarization that distinguishes a 
pyroelectric from a ferroelectric \cite{Scott00}.        
Inversion but not time-reversal symmetry 
is broken at a ferroelectric transition.  The development of a 
spontaneous polarization results from electric dipoles that are classical
and non-relativistic; they are spatially extended within the unit
cell. A ferroelectric displays a polarization-electric field hysteresis 
that is analogous to the magnetization-magnetic 
field hysteresis measured in magnetic materials.  Because the 
polarization is the electric dipole moment per unit volume it has the units of charge/area \cite{Lines77}.  Only the relative polarization, 
not its absolute value, is measured and this is usually performed by 
integrating a switching current \cite{Rabe07}.

\fight=4in
\fg{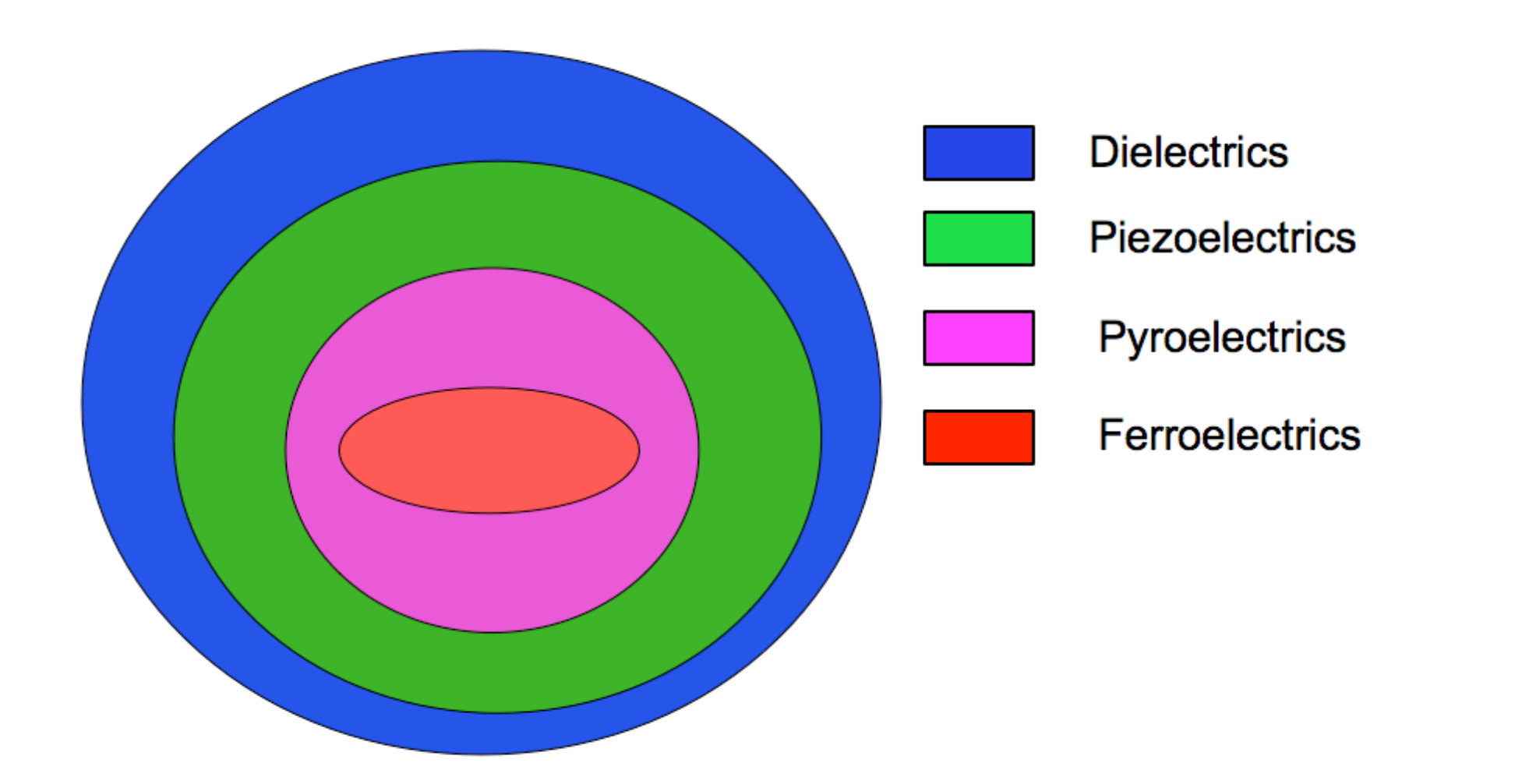}{fig6}{Venn diagram indicating graphically the relationship 
beween ferroelectrics, pyroelectrics, piezoelectrics and dielectrics.
Applied stress and temperature changes lead to electrical polarization in 
piezoelectrics and in pyroelectrics 
respectively \cite{Lines77,Fatuzzo67,Jona93,Scott00}; 
the switchability of this polarization in a field of practical
magnitude (and is less than the breakdown electric field) is what 
distinguishes a ferroelectric from a pyroelectric \cite{Scott00}.}

Qualitively there are two types of ferroelectric transitions \cite{Lines77}:
those driven mainly by amplitude fluctuations (displacive) and those 
driven mainly by angular fluctuations (order-disorder) at atomic scales.
In the latter case, the entropy change at the transition is higher than in 
the former situation.  At low temperatures, particularly as $T \rightarrow 0$,  
ferroelectic transitions are predominantly displacive and we'll return 
to this topic when we discuss
analogies with itinerant magnetism in the next section.
Here we are implicitly discussing ionic ferroelectricity where the 
polarization results from ionic displacements, though we do note
``electronic ferroelectricity'' in molecular 
crystals where the polarization is due to the ordering of 
electrons \cite{Kobayashi12}.  We emphasize that
ionic ferroelectrics can be order-disorder and/or displacive
in their character.  In these ferroelectics, strong coupling of the 
polarization and the lattice often leads to first-order transitions, 
both of order-disorder and displacive varieties.  

In conventional (ionic) ferroelectrics, the electric dipoles associated 
with the spontaneous polarization are produced by atomic rearrangements and they develop long-range order
at a ferroelectric transition.
Indeed the soft-mode theory of 
ferroelectricity \cite{Lines77,Anderson60,Cochran60,Scott74}, a lattice 
dynamics description, links the diverging dielectric response with
a vanishing phonon frequency and can indeed be viewed as an early model of
quantum criticality.  We can glean a flavor for the
soft-mode
approach by considering
the frequency-dependent electrical permittivity, $\epsilon_\omega$ 
of a simple diatomic 
harmonic lattice
\begin{equation}
\label{osc}
\epsilon_\omega = \epsilon_\infty + \frac{\epsilon_0 - \epsilon_\infty}
{1 - \frac{\omega^2}{\omega_{TO}^2}}
\end{equation} 
where $\epsilon_0$ and $\epsilon_\infty$ refer to the permittivities
at zero (static) and infinite frequencies respectively.
In the absence of free charge, the zero and the pole of 
$\epsilon_\omega$, respectively, determine the longitudinal and transverse optical mode 
frequencies $\omega_{LO}$ and $\omega_{TO}$ resulting in the
relation \cite{Kittel96,Ashcroft76}
\begin{equation}
\label{LST}
\frac{\epsilon_0}{\epsilon_\infty} = \left( \frac{\omega_{LO}}{\omega_{TO}} \right)^2
\end{equation}
that links the softening of a polar (transverse optical) phonon to the 
development of ferroelectricity.

This minimalist approach to soft-mode theory can of course
be generalized to include anharmonicities and many polar modes
where the frequencies are either measured \cite{Scott74} or 
calculated using first-principles 
methods \cite{Rabe07b,Cockayne00,Cockayne03}. 
We emphasize that a finite spontaneous polarization can only exist 
in a crystal with a polar space group \cite{Rabe07b}, 
though this does not ensure 
its switchability in a practical electrical field.  A structural 
signature of ionic ferroelectricity is that the finite polarization 
crystalline configurations result from small polar distortions of a 
high-symmetry (paraelectric) structure so that there is a simple 
pathway between them \cite{Rabe07b}.  In Figure \ref{fig7} we display the 
crystal structure of the well-studied perovskite ferroelectric $BaTiO_3$, its 
paraelectric (cubic) structure and two of its polarization states. 
From a first-principles perspective, a fingerprint of ferroelectricity is
the presence of unstable polar phonons in high-symmetry reference structures 
and this has been a successful method for characterizing known
and new ferroelectric materials \cite{Rabe07b}.
%Why does this work so well even though FE transitions are usually
%a mix of displacive and order-disorder?  Same story for the
%dielectric response?
%This issue should be discussed since we've just talked about
%displacive and order-disorder transitions. 
%Is this better to discuss this later when we talk specifically about STO?
%Should probably put this later in STO section. Interesting enough, 
%even though classical ferroelectrics are not purely displacive and sometimes 
%even display glassiness, their dielectric reponses are empirically dominated 
%by low-frequency zone-center optical phonons.
%Is there a simple way of understanding why this is true?
%this seems irrelevant; furthermore their presence 
%indicates that ferroelectrics are very sensitive to electromechanical 
%boundary conditions. 
Until relatively recently, it has been tacitly 
assumed that the polar phonon frequency vanishes as a function of 
temperature but of course other tuning parameters (like pressure) could 
achieve this softening as well.       

%Do we want to show the STO plot? Tricky since STO is not actually FE ! May
%be better when we discuss the material itself.

\fight=5in
\fg{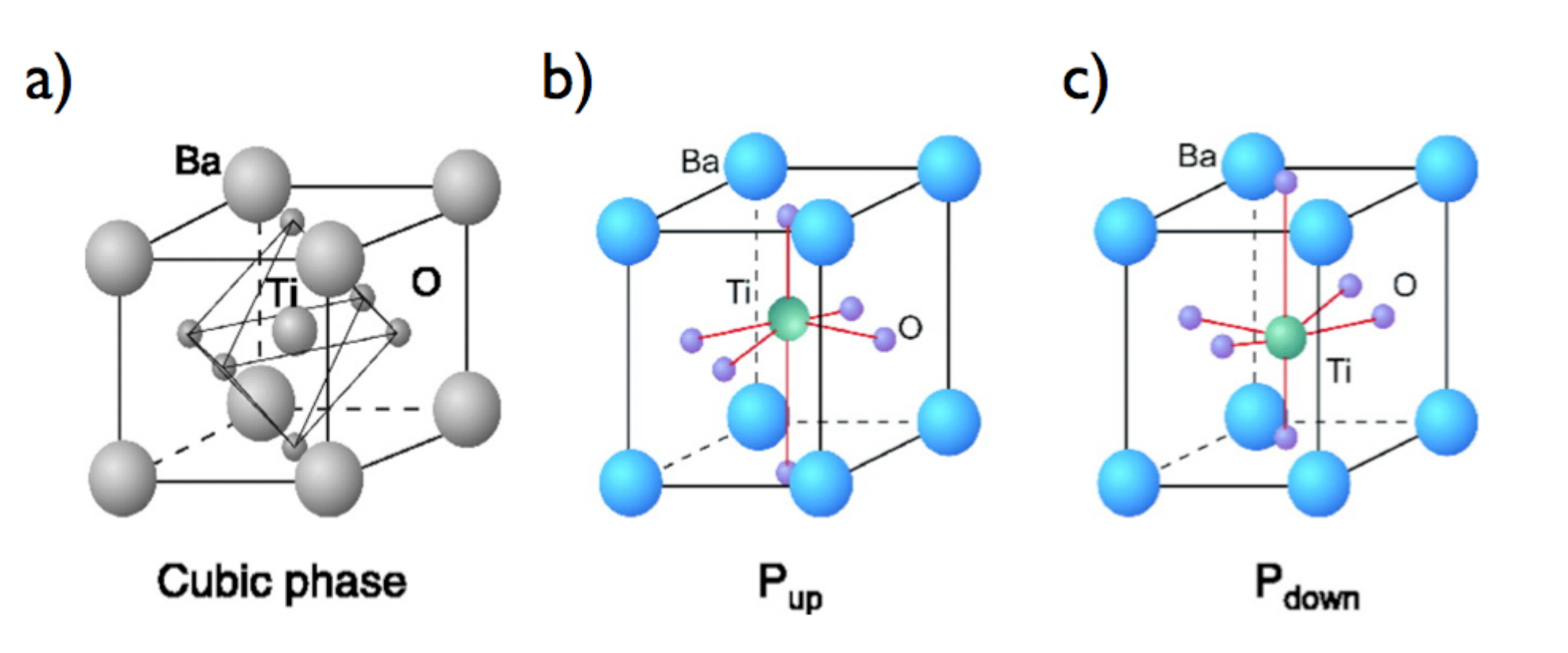}{fig7}{Crystal structures of the perovskite 
ferroelectric $BaTiO_3$. {\bf a)} High-temperature cubic paraelectric and 
room-temperature tetragonal ferroelectric structures for 
{\bf (b)} 
up and 
{\bf (c)} 
down polarizations respectively ($P_{up}$ and $P_{down}$) indicating the relative displacements of the positively charged Ti and negatively charged O ions; reprinted from Ahn et al. \cite{Ahn04} with permission.}

It is worth comparing the relative strengths of the electric
and magnetic dipole forces.  In atomic units $F_M$, the force between two magnetic dipoles at a distance $r$, is
\begin{equation}
F_M = \frac{\mu_0 \mu_B}{4 \pi r^3} \equiv \frac{\alpha_F^2}{4\pi} \left(\frac{a_B}{r}\right)^3
\end{equation}
where $a_B = 0.05 \ nm$ and $\alpha_F \equiv \frac{1}{137}$ are the Bohr 
magneton and the fine structure constant respectively; by contrast,
for an electric dipole $p = e \Delta a_B$, the dipolar 
interaction force is
\begin{equation}
F_D = \frac{p^2}{8 \pi \epsilon_0 r^3} \equiv \frac{\Delta^2}{4\pi} \left(\frac{a_B}{r}\right)^3,
\end{equation}
where the parameter $\Delta=O(1)$ is determined by effective charges and 
atomic 
displacements \cite{Born54}. The ratio of the ferroelectric to ferromagnetic 
dipolar forces is then of 
order $\left(\frac{\Delta}{\alpha}\right)^2 \equiv (137)^2$, 
indicating that long-range interactions are more significant 
in ferroelectrics than in generic magnetic systems.  This ratio
is a contributing factor towards explaining why the Ginzburg 
regime, where long-wavelength (``infrared'') fluctuations
govern the critical behavior, in ferroelectrics is empirically 
significantly smaller than its counterpart in magnets in many 
cases \cite{Chandra07};
classically the Ginzburg regime below $T_c$ is defined by the temperature 
interval close to a phase transition where order parameter 
fluctuations are comparable or larger than  
the average value of the order parameter itself.  However corrections to 
simple mean-field (Landau) theory due to 
anisotropic dipolar forces and anisotropic elastic interactions 
may be important.
For example, the first logarithmic corrections to mean-field exponents 
due to fluctuations at marginal dimensionality were calculated for a three-dimensional uniaxial 
ferroelectric \cite{Larkin69,Fisher73,Aharony76}; these predictions
were confirmed by experiment \cite{Ahlers75,Sandvold83} and played an 
important role in the development of the renormalization group 
approach to classical phase transitions \cite{Domb96,Fisher98}.
 
In the previous section we related $\langle \delta \phi_T^2 \rangle$ to 
$\chi(T)$ using (\ref{cfdt}), (\ref{qfdt}) and (\ref{chi}); let's now
apply these results to $d=3$ ferroelectrics where 
we are considering a QCP where the gap in the polar optical mode 
vanishes with a resulting dispersion
as $\omega \propto q$ as measured by neutron 
scattering \cite{Shirane67,Yamada69,Shirane74}
so the dynamical exponent $z=1$.    
In the proximity of a transition where $\alpha = 0$, we have
at long wavelengths ($q \rightarrow 0$)
\begin{equation}
\chi(T)^{-1} \approx T \int_\kappa ^{q_c} \frac{q^{d-1} dq}{q^2}
\label{chiequation} 
\end{equation}
where $q_c$ is the cutoff appropriate for the temperatures of interest;
here we are implicitly neglecting the temperature-dependence of $\kappa$ which,
according to (\ref{kappa2}), is reasonable for $T \rightarrow 0$ if 
$d + z > 4$.
At high temperatures ($T \gg \omega_q \ {\rm for} \ q < q_{BZ}$), $q_c = q_{BZ}$ 
has no temperature-dependence so we recover the Curie 
result $\chi^{-1} \propto T$; here we have assumed that $\kappa$ has 
saturated and thus is constant for these high temperatures.
However when quantum effects become important 
($q_T << q_{BZ}$), 
$q_c = q_T \propto T^{1/z}$; applying the results  
(\ref{kappa1}), (\ref{chiqt}) and (\ref{kappa2}) to $d=3$ 
ferroelectrics ($z=1$), we obtain
\begin{equation}
\chi^{-1} \propto T^{\frac{d-2 + z}{z}} \ = \ T^2
\label{chiqfe} 
\end{equation}
which we emphasize is distinct from the classical (Curie) behavior 
($\chi^{-1}\propto T$); since $d_{eff} = d + z = 4$ we also have log corrections 
that are usually difficult to observe experimentally. We 
note that we have reproduced a result first calculated 
diagrammatically \cite{Rechester71,Khmelnitskii73,Khmelnitskii71} 
and then rederived using other methods \cite{Roussev03,Palova09,Das09,Rowley14,Schneider76}.  

A critical reader may note that
here we have neglected the long-range dipolar interactions 
discussed previously; several 
theoretical studies \cite{Rechester71,Khmelnitskii73,Roussev03,Rowley14}
indicate that their main effect near a QCP is to 
produce a gap in the longitudinal fluctuations, but
that the transverse fluctuations remain critical.    
This conclusion is supported by recent measurements \cite{Rowley14}
of $\chi(T)$ near a ferroelectric QCP (FE-QCP) indicating 
good agreement with (\ref{chiqfe}). 
We should stress that at a QCP with $d + z > 4$, both $\kappa$ and
$q_T$ go to zero; however in this case, as we saw in (\ref{kappa2}), 
the ratio $\frac{q_T}{\kappa}$ diverges as $T \rightarrow 0$ so it
is the ``ultraviolet'' fluctuations that are crucial.  By contrast
at a classical transition,
$\kappa \rightarrow 0$ and the wavevector cutoff $q_c = \min\{q_T,q_{BZ}\}$ 
remains constant, and if $d < 4$ the ``infrared''
fluctuations are important.  The key roles of very different fluctuation 
regimes at classical and at quantum critical points suggests why the influence
of dipolar interactions is distinct in these two cases. 

Anisotropic elastic effects in ferroelectris have 
also been studied \cite{Brierley14}. The resulting domains have sufficiently 
slow dynamics, perhaps due to their physical extent or to pinning, that 
they do not seem to contribute to low-temperature thermodynamic quantities on 
measurable time-scales studied to date \cite{Rowley14}.

Analogous to Einstein's approach to the specific heat problem \cite{Kittel96}, 
we can also consider the
situation where the low-energy excitations are dispersion-free with a single 
frequency $\omega_0$.  This is just the case of a simple harmonic 
oscillator \cite{Coleman15} so
we have
\begin{equation}
\chi(\omega) \propto \frac{\omega_0}{\omega^2 - \omega_O^2}
\end{equation}
and
\begin{equation}
\chi^{''}(\omega) \propto \frac{\delta (\omega - \omega_0)}{\omega_0}.
\label{imsho}
\end{equation} 
Using the identity for the Bose function  
\begin{equation}
n\left(\frac{\omega}{T}\right) + \frac{1}{2} = \frac{1}{2} \coth \left(\frac{\omega}{2T}\right),
\label{Bidentity}
\end{equation}
we input (\ref{imsho}) into the general expression for the moment amplitude variance (\ref{ggfdt}) to obtain
\begin{equation}
\langle \delta \phi^2 \rangle \propto \frac{1}{\omega_0} \coth\left({\frac{\omega_0}{2T}}\right).
\label{Einsteinvar}
\end{equation}
Taking the $q \rightarrow 0$ limit of (\ref{Efluct0}) we obtain
\begin{equation}
\chi^{-1}  = (\alpha + 3 \beta \langle \delta \phi^2 \rangle) 
\label{chi0}
\end{equation}
where $\alpha$ and $\beta$ are defined in (\ref{fLandau}); 
both are finite since we are not at a phase transition.
Combining (\ref{Einsteinvar}) and (\ref{chi0}), we then obtain 
\begin{equation}
\chi^{-1} = \left[\alpha + \frac{3 A \beta}{\omega_0} \coth \left(\frac{\omega_0}{2T} \right) \right]
\end{equation}
which can be rewritten in the Barrett form \cite{Barrett52,Salje91}
\begin{equation}
\chi =  C \left[\frac{\omega_0}{2} \coth\left(\frac{\omega_0}{2T}\right) - T_0 \right]^{-1}
\label{barrett}
\end{equation}
where $C = \frac{\omega_0^2}{6\beta}$ and $T_0 = -\frac{\alpha A}{6\beta}$ 
are constants written in terms of the original parameters.  We 
re-emphasize that the Barrett (or rather ``Einstein-Barrett'') expression 
is for dispersion-free 
excitations \cite{Lines77}; it is thus not valid in the immediate 
vicinity of a quantum critical point where, similar to the situation in the
Debye model \cite{Kittel96,Ashcroft76}, 
excitations of different wavevectors have different frequencies.

The Gr\"uneisen ratio, $\Gamma = \frac{\tilde{\alpha}}{c_P}$ 
where $\tilde{\alpha}$ and $c_p$ are the thermal expansion and the 
specific heat respectively, has been identified
as a physical quantity that diverges at a QCP and is {\sl constant} at
a classical critical point \cite{Khmelnitskii71b,Khmelnitskii75,Zhu03}. 
The Gr\"uneisen ratio is then a 
useful bulk thermodynamic probe to locate, classify and categorize QCPs 
in a diverse set of materials, so let's now use the methods we've developed
to determine $\Gamma(T)$
near a FE-QCP.  As an aside, we note that this 
Gr\"uneisen ratio is to be distinguished from the Gr\"uneisen parameter 
that measures the logarithmic change of a particular mode frequency 
as a function of volume change; the two quantities are only simply related when 
the lattice frequencies are temperature-independent which is definitely 
not the case for the (predominantly) displacive ferroelectrics (DFEs) of 
interest here.

Using Maxwell's relations, the Gr\"uneisen ratio can be 
written as the effect of a volume change on a solid's total thermal
energy, $\Gamma = \frac{1}{V} \frac{\partial V}{\partial  U}$.  
Because $d=3$ displacive quantum paraelectrics ($z=1$) reside in their 
marginal dimension ($d_{eff} = 4$), their critical
behavior can be described by a self-consistent mean-field theory where
fluctuation corrections due to anharmonicities are included via
the fluctuation-dissipation theorem;
we've already implemented this approach in (\ref{Efluct0}) where the 
Gaussian fluctuations are treated 
to leading order using ({\ref{qfdt}). This approach is only strictly 
valid for $d_{eff}  > 4$, but the weakly temperature-dependent logarithmic
corrections to mean-field theory are likely to be too small to be observable in most experiments \cite{Rowley14}.  The free energy as a function of the polarization change
($\delta \phi_E$ where here $\phi_E$ is the electric dipole) 
\begin{equation}
F(\delta \phi_E,\delta V) = \frac{\alpha}{2} \delta \phi_E^2 + \frac{a}{2} \delta V^2
- \eta (\delta V)(\delta \phi_E^2)
\label{fenergy}
\end{equation}
where on symmetry grounds the form of the coupling term is even in 
$\delta \phi_E$ 
but odd in $\delta V$, the change in volume
from the equilibrium $T=0$ value; $\alpha=0$ at
a phase transition and  $a$ and
$\eta$ are constants.  

\vskip0.3in

\input KIR_Table1

\vskip0.25in

Minimizing (\ref{fenergy}) with respect to volume
and, using ({\ref{qfdt}) to average over fluctuations 
to get the most probable result, we obtain
\begin{equation}
\langle \delta V \rangle \propto \langle \delta \phi_E^2 \rangle 
\label{deltaV}
\end{equation}  
so that 
\begin{equation}
\Gamma_{FE} (T)  = \frac{1}{V} \left(\frac{\delta V}{\delta U}\right) 
\propto \frac{\langle \delta \phi_E^2 \rangle }{\delta U}. 
\label{gamma}
\end{equation}
Because neither the numerator or the denominator has a singularity in
$(T-T_c)$ for a finite transition temperature $T_c$, we expect that 
\begin{equation}
\Gamma_{CFE} (T \rightarrow T_c) \propto (T - T_c)^0 
\label{gammacfe}
\end{equation}
will be independent of temperature;
this is supported by 
experiment reporting the identical temperature-dependences of
thermal expansion and specific heat near finite-temperature ferroelectric phase 
transitions \cite{Lines77}. 

However in the approach to a $T \rightarrow 0^+$ FE-QCP, we can use
(\ref{chi}) to write
\begin{equation}
\lim_{T \rightarrow 0^+} \langle \delta \phi_E^2 \rangle \propto \chi^{-1}
\propto T^2.
\end{equation}
Analogous to the Debye approach to the specific heat \cite{Ashcroft76},
the change in energy is equal to the temperature multiplied by the number of
accessible modes
\begin{equation}
\delta U_{QFE}  \propto T (q_T^d)
\end{equation}
so that the temperature-dependence
of $\Gamma$ in the vicinity of a ($d = 3$) FE-QCP is
\begin{equation}
\Gamma_{QFE}  = \left(\frac{\delta V}{\delta U}\right) 
\propto
\left(\frac{\langle \delta \phi_E^2\rangle}{\delta U}\right)   
\propto \frac{\chi^{-1}}{T q_T^d} = \frac{T^2}{T^4} = \frac{1}{T^2}      
\label{gammaqfe}
\end{equation}
that {\sl diverges} with decreasing temperature and thus 
is dramatically different from the temperature-independent
classical case (\ref{gammacfe}); here we are implicitly considering
the strongly temperature-dependent part of $\phi_E$.

Since $\Gamma = \frac{\tilde{\alpha}}{c_P}$ where $\tilde{\alpha}$ and 
$c_P$ are the thermal expansion and the specific heat respectively,
its experimental determination involves two distinct measurements.  Not only 
does the temperature-dependence of $\Gamma$
signify the importance of quantum fluctuations, but it is also an independent 
determination \cite{Rowley16}  of the dynamical exponent $z$.  
In Table I. we summarize the distinctive temperature-dependences of the inverse susceptibility and the Gr\"uneisen ratio in the vicinity of three-dimensional classical and quantum displacive ferroelectric critical points.

\vskip0.2in

\input Sec3_Box_FE_Necessities

\vskip0.2in

More generally paraelectrics near displacive ferroelectric quantum 
critical points offer appealing examples of quantum critical behavior 
often without the complications of dissipation and damping that occur in 
metallic magnetic systems.  Furthermore because their dispersion 
is linear ($z=1$), quantum critical paraelectrics
can be studied just below, at or just above their upper critical  
dimension ($d^{upper} = 3 + 1 = 4$) making detailed comparison between 
theory and experiment possible in ways that are not so straightforward for 
their metallic magnetic counterparts (e.g. $z=3$ for a metallic 
ferromagnet)\cite{Sachdev99,Roussev03,Rowley14}. It is thus 
perhaps not so surprising that some of the earliest 
theoretical studies of quantum criticality were done in a paraelectric 
setting \cite{Rechester71,Khmelnitskii73}.

A key similarity between displacive ferroelectrics (DFEs) and metallic magnetic
systems is that in both material classes amplitude fluctuations of the 
appropriate moments on length-scales of order their unit cells
are significant 
so that it is relatively straightforward to suppress their orderings to 
$T \rightarrow 0$.   By contrast, in insulating magnets and order-disorder 
ferroelectrics the moment fluctuations are mainly orientational
on length-scales of order their unit cells 
in the high-temperature phase; it is therefore challenging to
prevent ordering at low temperatures
for the study of quantum criticality, though there are indeed some 
magnetic examples \cite{Sachdev08,Batista07,Jaime04,Giamarchi08,Kraemer12}.
As an aside, we should note that in the literature the descriptives metallic
and itinerant are often used interchangeably; here we will use both
terms to mean that the volume of the Fermi surface encloses the 
magnetic carriers. Of course the dynamics in displacive ferroelectrics 
(propagating vibrational modes) are distinct from those in itinerant magnets 
(spin precession and dissipative spin dynamics) and this will 
result in different quantum critical behavior.
The issue of universality near quantum phase transitions is still
one of open discussion, and a new class of materials for detailed
study could shed light on this central issue \cite{Canfield08}.
%We emphasize that in coarse-grained models both classes of materials will
%have amplitude and orientation fluctuations in their respective moments; this
%is indeed a key reason why critical phenonemena is expected to be universal
%though specifics associated with universality classes are still being worked 
%out by the community.  
With this goal in mind, in Table II.  we summarize 
key similarities and 
differences between displacive ferroelectrics and itinerant ferromagnets, 
focussing on characteristics most relevant for the study of 
quantum criticality.

%\vskip0.15in

\input KIR_Table2

\vskip0.1in

%% file: KIR_Table1
%\begin{tcolorbox}

\begin{center}

\begin{tabular}{|c| c| c |}
\hline
\hline
& & \\
& {\bf $T \rightarrow T_c^+$ \quad $(T_c > 0)$ } & {\bf $T \rightarrow 0^+$}
\quad $(T_c = 0)$ \\ & &  \\ \hline & & \\
{\bf Inverse Dielectric} & & \\ 
{\bf Susceptibility} & $T$  & $T^2$  \\ 
 $\chi^{-1}$ &  &  \\  & & \\
\hline & & \\ 
{\bf Gr\"{u}neisen Ratio} & &\\ & & \\
$\Gamma = \frac{\alpha}{c_P}$ & $T$  & $T^{-2}$  \\
 &  &   \\
& {\bf Constant} & {\bf Diverging} \\ & & \\
\hline
\end{tabular}

\end{center}

\vskip0.05in

Table I:  Expected temperature-dependences of two experimental probes in the 
approach to $d=3$ ferroelectric critical points %using the self-consistent mean-field approach discussed in the text, where 
we reproduce susceptibility results found elsewhere 
\cite{Rechester71,Khmelnitskii73,Roussev03,Palova09,Das09,Rowley14,Khmelnitskii71, Schneider76}.  
Here $T \rightarrow T_c^+$ and $T \rightarrow 0^+$ refer to classical and to 
quantum critical points respectively.  In the approach to a classical 
critical point, the inverse dielectric susceptibility displays Curie 
($\chi^{-1} \propto T$) behavior; for $T \rightarrow 0^+$, it scales 
as $\chi^{-1} \propto T^2$ 
where here we are neglecting weak logarithmic corrections for the 
relevant case $d + z =4$.  We note that $\chi = \epsilon - 1$ 
where $\epsilon$ is the dielectric function.  
The Gr\"{u}neisen ratio, $\Gamma = \frac{\alpha}{c_P}$ 
where $\alpha$ and $c_p$ are the thermal expansion and the 
specific heat respectively, diverges near a quantum critical point 
($\Gamma \propto T^{-2})$; by contrast it remains
constant near a classical one and thus is an important signature of quantum 
criticality \cite{Khmelnitskii71b,Khmelnitskii75,Zhu03}.

%\end{tcolorbox}

%% file: Sec3_Box_FE_Necessities
\begin{tcolorbox}

\center{\bf Ferroelectric Necessities:  Key Concepts}
\vskip0.20in
\begin{itemize}
\vskip0.10in

\item A ferroelectric has a {\bf spontaneous polarization} that is {\bf switchable} by an electric field.

\item {\bf Inversion} symmetry is broken in the ferroelectric phase.

\item The {\bf temperature-dependence of observable quantities} (e.g. susceptbility) in the vicinity of both classical and quantum critical points can be determined using a {\bf self-consistent mean-field theory where fluctuation corrections due to anharmonicities are given by the fluctuation-dissipation theorem}.

\item {\bf The Barrett form of} $\chi(T)$ results if {\bf a single Einstein frequency} is assumed; this is {\bf not valid} in the {\bf vicinity of a QCP} where the wavevector-dependence of the excitation spectrum (dispersion) is important.

\item The {\bf Gr\"uneisen ratio diverges} with decreasing temperature {\bf near a quantum ferroelectric critical point} but remains constant near its classical counterpart.

\end{itemize}

\end{tcolorbox}

%% file: KIR_Table2
%\begin{tcolorbox}

\begin{center}

\begin{tabular}{|c| c| c |}
\hline
\hline
   & {\bf Displacive Ferroelectrics} & {\bf Metallic Ferromagnets} \\ \hline
\multirow{5}{*}{\bf Dipole Origin} & Charge Separation  & Bohr Magnetron of Electron\\ 
 & & (and Possible Orbital Motion) \\
 & Classical & Quantum \\
 & Non-Relativistic & Relativistic \\ 
 & No Intrinsic Angular Momentum & Intrinsic Spin Angular Momentum \\ 
\hline
\multirow{2}{*}{\bf $T > T_c$} & \multicolumn{2}{|c|}{Dipole Moments Ill-Defined Due to Amplitude Moment Fluctuations} \\ 
  & \multicolumn{2}{|c|} {Moment Fluctuation Energy Scale \ $> T_c$} \\ \hline
\multirow{3}{*} {\bf Dynamics} & Propagating & Precessional and Dissipative\\
& Atomic Vibrations  & Spin Fluctuations \\
 & (Second-Order in Time) & (First-Order in Time) \\ \hline
{\bf Dynamical Exponent z} & 1  & 3 \\ 
{\bf $(\omega \propto q^z)$} &  & (Assuming Landau damping) \\ \hline
{\bf $d^{upper}_{space} = 4 - z$} & 3  & 1\\
\hline
\end{tabular}

\end{center}

\vskip0.05in

Table II:  Key Similarities/Differences between Displacive Ferroelectrics and Metallic Ferromagnets Most Relevant for the Study of Quantum Criticality. 

\vskip0.1in

%\end{tcolorbox}

%% file: Sec4_STO.tex
%8/7/17 
%
%The Case of $SrTiO_3$ (STO) to Date

So far we've discussed quantum criticality in displacive ferroelectrics
in rather broad, abstract terms...let's now turn to what all this means
specifically for the case of $SrTiO_3$ (STO), a material that has been an
important setting for basic research and for specific applications over 
the course of several decades \cite{Lines77,Lemanov02}.  Here we will
focus mainly on summarizing its low-temperature properties, where more
detail can be found in reviews (and references therein) 
elsewhere \cite{Samara01,Kvyatkovskii01,Lines77,Lemanov02,Dec05}.  As we have
already discussed, ferroelectricity in the $ABO_3$ perovskites is 
driven predominantly by soft long-wavelength transverse optical (TO) phonons; 
thus this displacive ferroelectric (DFE) phase transition is naturally 
sensitive to pressure-tuning and hence to studies of quantum criticality.
$BaTiO_3$ (BTO) was the first perovskite ferroelectric to be identified,
and the development of FE from its simple high-temperature cubic perovskite
structure was very appealing and led to intense study \cite{Lines77}.  
At high temperatures, the dielectric response of $SrTiO_3$, an isovalent
cousin of BTO, is Curie-Weiss and suggests a ferroelectric temperature
of $T_c \sim 40 K$.  Like BTO, STO has a soft TO mode such that
$\epsilon^{-1} \propto \omega^2$ over a broad temperature region \cite{Scott74}.
%this might make a nice figure to show what a nice DFE this is!
However at $T_c = 105 K$, STO has a cubic-tetragonal (C-T) transition
where both phases are paraelectric in contrast to the C-T transition in BTO
where FE develops. In STO there are clear thermodynamic anomalies at $T_c$ but
no inversion symmetry-breaking, though at low temperatures boundaries
between tetragonal domains are polar \cite{Salje13,Ma16}. Phonon softening at the
Brillouin zone boundary is observed at $T_c$ and this 
antiferrodistortive (AFD) transition in STO is associated with the 
development of staggered rotations of oxygen octahedra in adjacent unit 
cells.  Though STO polar soft modes are present, ferroelectricity is not 
observed to the lowest temperatures measured at ambient pressure \cite{Rowley14}.

\fight=5in
\fg{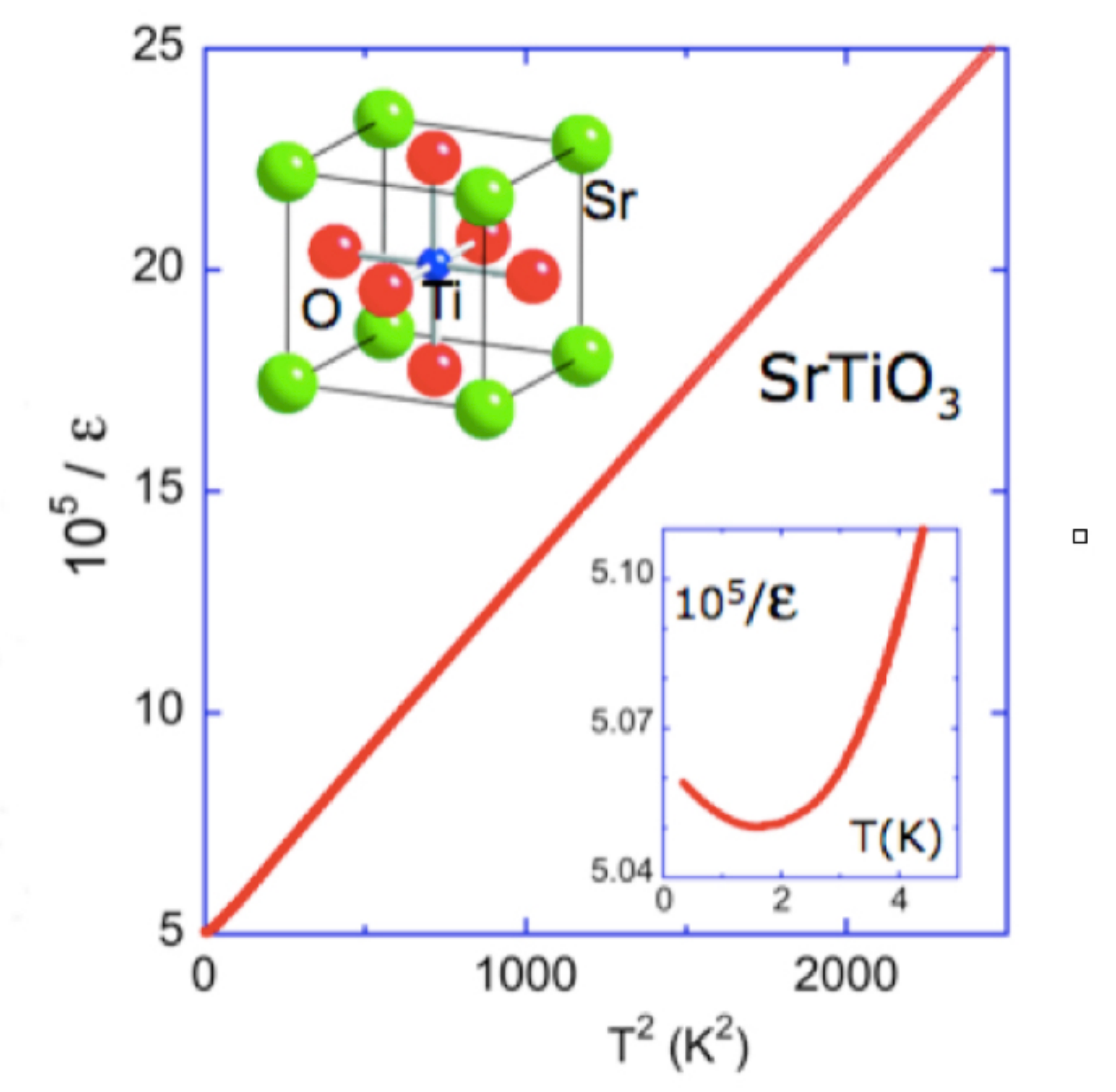}{fig9}{Temperature-dependance of the inverse dielectric
function $\epsilon^{-1} (T)$ at ambient pressure for $SrTiO_3$  
as a function of the square of 
the temperature up to approximately $T = 50 K$
from \cite{Rowley14} indicating good agreement with the
behavior $\epsilon^{-1} \propto T^2$ expected 
theoretically ($\epsilon = 1 + \chi$) in the approach to a 
$d=3$ ferroelectric quantum critical 
point where the 
weak logarithmic corrections are not observed \cite{Rechester71,Khmelnitskii73,Roussev03,Palova09,Das09,Rowley14,Khmelnitskii71,Schneider76}.   
The room-temperature cubic perovskite crystal structure of $SrTiO_3$ is 
shown in the top left corner.
The lower inset is an 
expanded view of the low-temperature data \cite{Rowley14}, 
indicating an upturn below $4 K$ most likely due to 
coupling of the polarization 
with acoustic phonons \cite{Palova09,Rowley14,Khmelnitskii71,Rowley14b,Hayward05}.
}

The unexpected low-temperature behavior in the dielectric 
response of STO (it is large but finite as shown in Figure \ref{fig9}) 
led to STO being named the first 
``quantum paraelectric'' \cite{Muller79}.  It was assumed
that the stability of the paraelectric state in low temperature STO is due
to effects of zero-point fluctuations analogous to the situation
in liquid helium where crystallization is never achieved at ambient 
pressure. There was already prior theoretical literature on 
the effects of quantum fluctuations on low temperature displacive 
transitions \cite{Rechester71,Khmelnitskii73,Khmelnitskii71,Schneider76,Barrett52}, and experiments on STO stimulated more theoretical research in 
this direction \cite{Kvyatkovskii01,Roussev03,Palova09,Das09,Rowley14,Martonak94,Zhong96,Conduit10}.  Usually one associates zero-point fluctuations
with light atoms like hydrogen or helium so their significance for STO
may seem surprising.  However quantum effects can also assume importance
when there are two or more low-temperature phases present, for example 
paraelectricity and ferroelectricity,  
with negligible energy differences \cite{Lines77}.  In the case of STO,
the coupling between the oxygen rotations and the soft polar mode is very
small so that quantum fluctuations can affect the AFD and the FE effectively
independently \cite{Hayward05};
computationally quantum fluctuations have been 
shown to suppress the FE transition 
\cite{Zhong96}, supporting the 
proposal that STO is a quantum paraelectric. 
It was noted early on that the Einstein-Barrett 
expression (\ref{barrett}) \cite{Barrett52} 
for the dielectric susceptibility does not work well for STO \cite{Muller79},
most likely because STO has a phonon dispersion \cite{Lines77}.
Indeed it is exactly why STO is of interest to us at low temperatures
since we expect scale-free quantum fluctuations there to be quite important.

The antiferrodistortive transition in STO at $T_c = 105K$ at ambient
pressure is very close to a tricritical point and indeed STO is
a marginal system very close to the stability edge of its paraelectric
phase.  External perturbations including uniaxial stress, epitaxial strain
and chemical subsitution induce ferroelectricity
at finite temperatures.  More recently it has been 
found \cite{Kvyatkovskii01,Itoh00,Wang01,Scott11b}
that ferroelectricity
can also be induced in STO with isotope subsitution (Oxygen-18) such
that for $SrTi(^{16}O_{1-x}^{18}O_x)_3$ the ferroelectric transition
temperature scales as   $T_{FE} \propto (x - x_c)^{0.5}$ for
$x \ge x_c \approx 0.3$ where $T_{FE} = 23 K$ for $x=1$.  
In the simplest models isotope subsitution softens the polar phonons, 
and there are several 
such theoretical discussions specific to STO 
\cite{Kvyatkovskii01,Bussman-Holder00,Itoh04,Bussmann-Holder08}; here the key assumption
is that the mass increases at constant stiffness.  However we 
might also expect that a decrease in frequency increases the 
susceptibility and thus decreases the stiffness, leading to an 
increase in fluctuation amplitude.  The relative importance of mass vs. 
stiffness change in describing isotopic substitution in STO is a topic 
of current discussion. 

On the experimental side, application of hydrostatic pressure to STO-18 ($x=1$) suppresses
its ferroelectric transition to zero-temperature \cite{Venturini04},
so that the effects of quantum fluctuations can be studied precisely at the
QCP.
More recently the dielectric response of $SrTi(^{18}O_x^{16}O_{1-x})_3$ 
has been studied for varying $x$ at very low temperatures at ambient pressure; 
because it does not depend strongly on sample growth conditions or purity, 
it has
been suggested that disorder is not a key feature \cite{Rowley14}.
The detailed behavior of the dielectric response is in excellent
agreement with theoretical 
predictions \cite{Kvyatkovskii01,Khmelnitskii73,Roussev03,Palova09,Das09,Rowley14,Khmelnitskii71,Schneider76,Rowley14b,Conduit10}, suggesting
that this is a system where detailed interaction between theory and
experiment are possible.  Work is currently in progress on the Gr\"uneisen
ratio \cite{Zhu03} in this same set of materials to explore its 
behavior at and in
proximity to the DFE-QCP (displacive ferroelectric quantum critical 
point) \cite{Rowley16}.  We note it is
necessary to take account of the coupling of the electronic polarization field 
with the acoustic phonons to obtain a full description of the dielectric
behavior particularly at the very lowest temperatures, below a few 
Kelvin \cite{Palova09,Rowley14,Rowley14b,Hayward05}.

\fight=5in
\fg{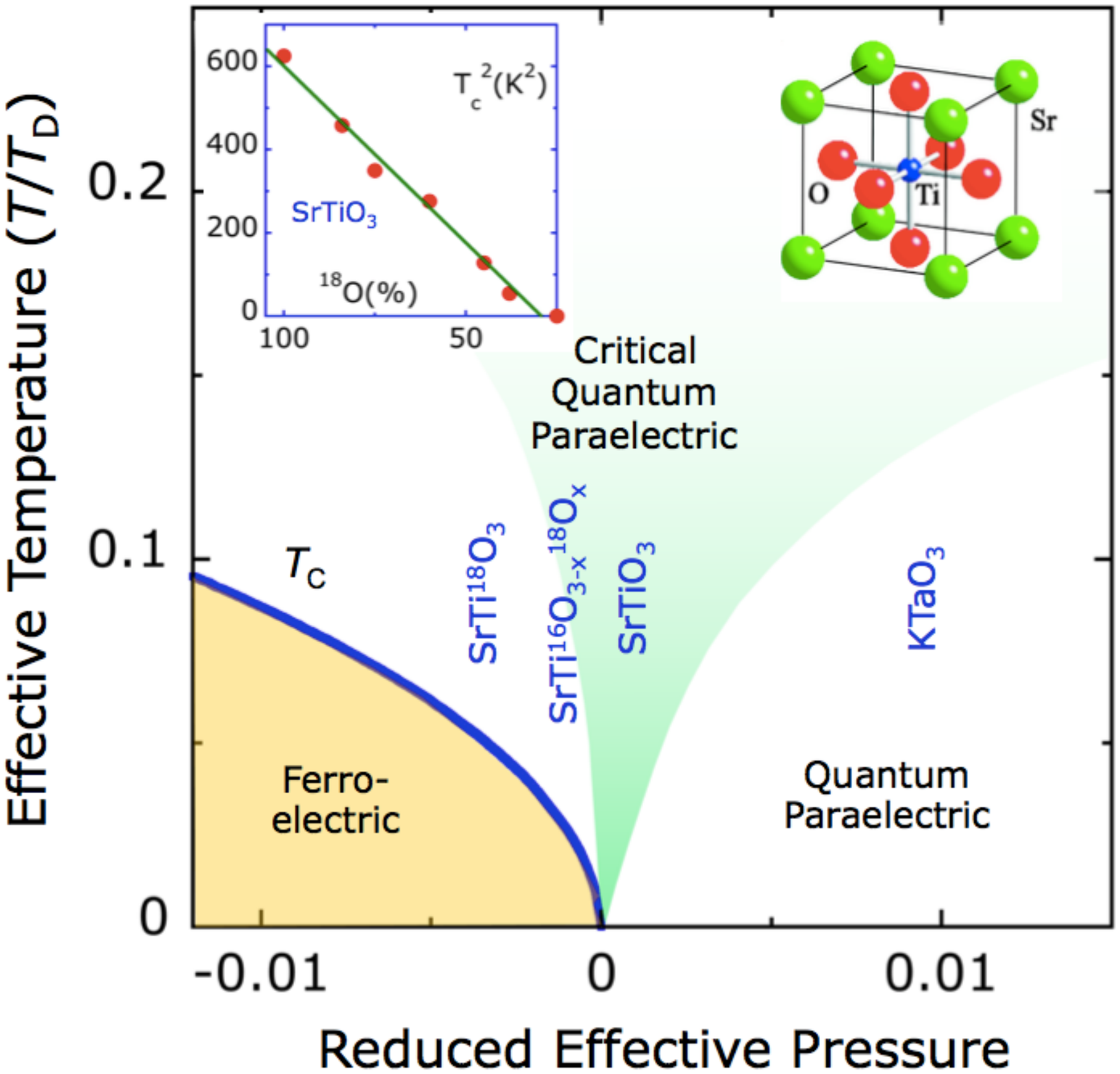}{fig10}{Effective temperature vs. reduced effective 
pressure phase diagram for 
$SrTiO_3$, $KTaO_3$ and related materials.  Here the effective temperature
is the ratio of the temperature and the material's Debye temperature associated
with its optical phonon branch
$\left(\frac{T}{T_D}\right)$.
The effective pressure can be tuned  by isotopic ($SrTi(^{18}O_x^{16}O_{1-x})_3$)
or by chemical ($Sr_{1-x}Ca_xTiO_3$) substitution \cite{Bednorz84,Rischau17}, or by application of external 
hydrostatic pressure \cite{Venturini04}.
Based on an integrated theoretical-experimental approach \cite{Rowley14}, 
a selection of  materials is positioned on this phase diagram (with units of effective pressure
defined in \cite{Rowley14}) where a critical quantum paraelectric is one 
with a gapless dispersion ($\omega \propto q^z$) whereas the Einstein-Barrett 
description \cite{Barrett52} may only apply to 
materials in the ``quantum paraelectric'' phase with a gapped spectrum.
Insert:  $T_c^2$ vs $^{18}O$ percentage in
$SrTi(^{18}O_x^{16}O_{1-x})_3$ with a linear slope indicating an 
isotopically-tuned ferroelectric phase transition temperature 
with $T_c \propto \sqrt{x}$, a result in agreement with
self-consistent mean-field theory \cite{Rowley14}.  The room-temperature cubic perovskite structure of $SrTiO3$ is also shown in the top of 
the phase diagram.  
This figure is adapted from Rowley et al. \cite{Rowley14}.}

For the sake of completeness, we should add that although the transverse optic soft mode in $SrTiO_3$ reaches zero frequency only in STO-18
causing ferroelectricity below roughly $30 K$, there is a different and rather unexpected kind of short-range ferroelectric distortion in all isotopic variations of STO:  below roughly $80 K$, the $Sr$-ions displace along $[111]$ directions, yielding a triclinic structures with local polarizations \cite{Salje13,Ma16}.  Under normal conditions, these local polarization cannot all be aligned to yield a macroscopic polarization, so in some important way cryogenic STO with $^{18}O$
does not behave as a conventional 
paraelectric. The ferroelectric nanodomains are nestled inside larger ferroelastic domains (``walls within walls'') \cite{Salje13}. This local symmetry 
may play a role  in the crystallographic structure of ferroelectric STO 
with $^{18}O$, and this remains an open question.  Again we note that 
the response time of these domains appears to be very slow \cite{Brierley14}
as they don't appear to contribute to observed low temperature thermodynamic quantities studied so far \cite{Rowley14}.

In a nutshell, STO and its isotope variants, provide a nice setting
to study quantum criticality since the dynamics are simple (propagating)
and it resides at its upper critical spatial dimension 
$d^{upper}_{space}  = 4 - 1 = 3$  so that results from 
both scaling and self-consistent phonon theories apply (up to 
logarithmic corrections) and can be compared in detail with experiment.
In Figure \ref{fig10} we display a schematic Temperature-Pressure phase diagram
indicating the observed behavior of $SrTiO_3$ and related perovskite materials
at ambient pressure.
Of course there are a number of other exciting recent developments
associated with STO at low temperatures that also present exciting
research opportunities both for fundamental study and also towards applications,
and we mention them briefly here:

\begin{itemize}

\item {\bf Giant Piezoelectricity.}
The large piezoelectric response of STO at low temperatures makes it
very useful for a number of cryogenic applications \cite{Grupp97}.  To our 
knowledge, the piezoelectricity of the isotopically mixed STO family has not 
been systematically measured and it may be tunable as a function of the 
$^{18}O$/$^{16}O$ ratios and epitaxial strain to suit specific needs.

\item {\bf Photoinduced Enhanced Dielectric Constant.}
It has been found that a significantly
enhanced dielectric constant can be induced in 
STO by ultraviolet radiation with
the suggestion that it is related to quantum effects \cite{Takesada03,Hasegawa03}, possibly through large polaron formation \cite{Nasu03}

\item {\bf Chemically Doped STO.}
There has been extensive work on the low temperature
properties of chemically doped quantum paraelectrics \cite{Kleeman00},
particularly on impurity-induced ferroelectricity.  The development
of quantum relaxors and quantum paraelectric glassiness has been
less studied and could be important \cite{Ang98}, as we'll discuss in
the next section, for electrocaloric applications. 

\item{\bf Electron Transport in Doped STO.}
Electron transport in n-doped $SrTiO_3$, achieved either
by oxygen reduction or by Nb subsitution, has been 
observed \cite{Spinelli10}, with high carrier mobility \cite{Son10,Behnia15} 
and unusual resistive behavior \cite{Lin15}.
The magetoresistance and the Hall resistivity associated
with photoinduced carriers
in STO is also unconventional \cite{Kozuka08} suggesting
that the metallic state emerging from doped STO may need further
characterization particularly due to its very low Fermi temperature.

\item {\bf Superconductivity in STO.}
Electron-doped STO is one of the most dilute 
superconductors known \cite{Lin13,Prakash16},
and most likely a non-BCS mechanism is necessary for its explanation 
More recently a gate-tunable insulating-superconducting transition has
been observed in an STO weak link \cite{Gallagher14}, again pointing
to anomalous behavior in this material.  The
dependence of the superconducting $T_c$ on the percentage of $^{18}0$ in 
the STO is an active topic of theoretical \cite{Edge15,Kedem16} and 
experimental \cite{Stucky16} research.
We will return to the question of superconductivity in STO 
in the ``Open Questions'' section.

\end{itemize}

These are just some of the many stimulating questions associated with STO
at low temperatures.  Of course this material is very much in the
news at higher temperateratures including its role in 
oxide interfaces \cite{Ohtomo04,Lerer11,Sulpizio14}
and as a substrate that mysteriously enhances the superconductivity in 
$FeSe$  \cite{Ge15}. 

In this section we have focussed on ferroelectric quantum criticality
in STO, and we conclude it by noting that ferroelectric quantum
phase transitions have been observed in a variety of systems 
including other insulating perovskites \cite{Ishidate97}, 
organic complexes \cite{Horiuchi03,Horiuchi15,Rowley15} and 
narrow-band semiconductors \cite{Suski83}.
In order to emphasize this point, in Figure \ref{figxx} we display 
four distinct examples of ferroelectric quantum transitions, noting the range of $T_c$'s accessible with chemical substitution and applied pressure.

\fight=5.0in
\fg{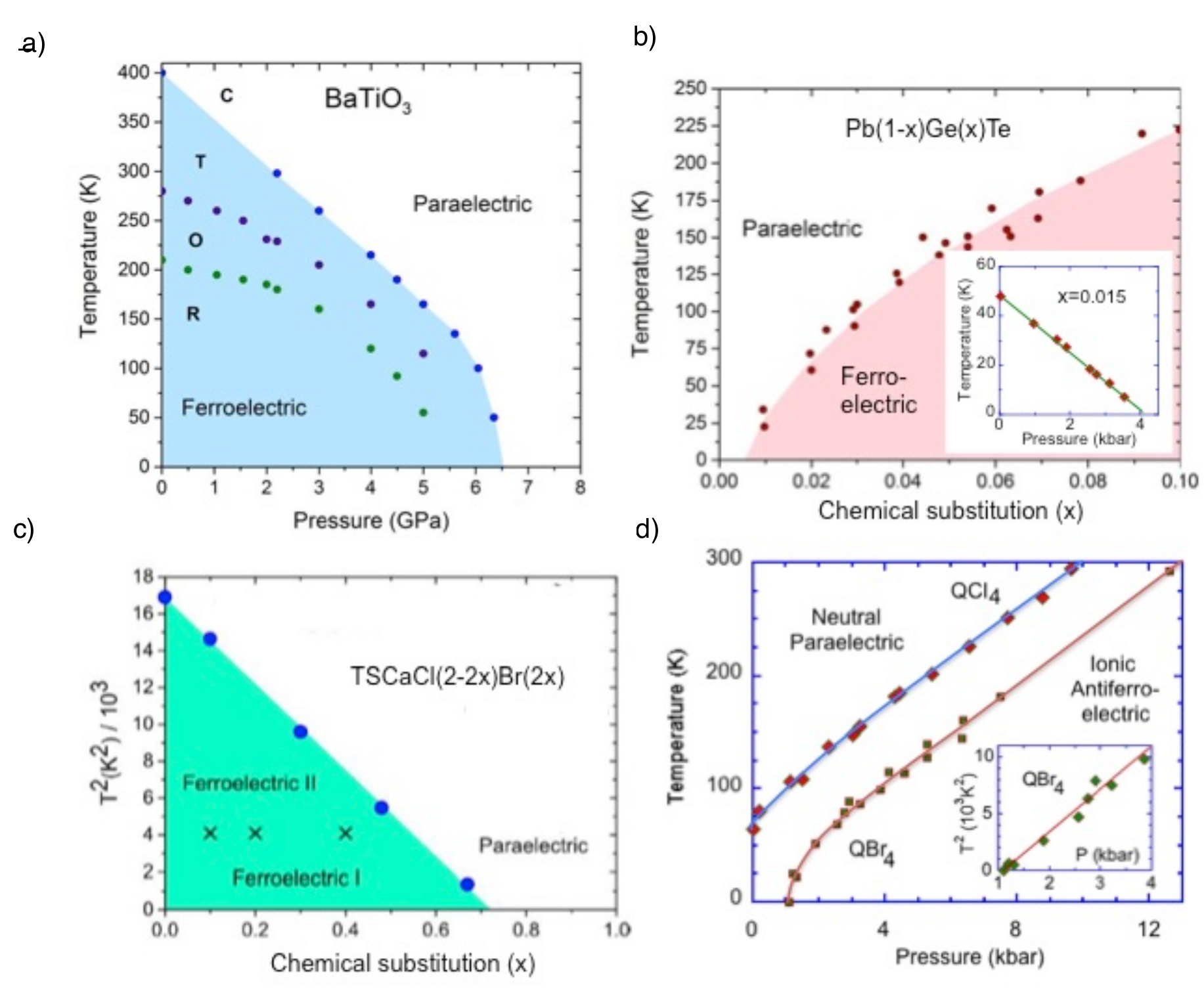}{figxx}{Four phase diagrams indicating 
different materials where ferroelectric quantum phase transitions have been
studied experimentally with tuning by pressure or by chemical substitution. 
{\bf a)} Pressure-tuned ferroelectric quantum phase transition in perovskite $BaTiO_3$. The figure labels C,T,O and R refer to the cubic, tetragonal, orthorhombic and rhombohedral structural phases of $BaTiO_3$.  The polarization direction points in different directions in each of the three ferroelectric phases (T, O and R).  All transitions are first-order at ambient pressure.  This figure is adapted from Ishidate et al. \cite{Ishidate97} with permission.  {\bf b)}  The IV-VI family of narrow-band semiconductors $GeTe$ and $PbTe$ have soft transverse-optical phonon modes that can lead to ferroelectric instabilities.  Pressure, carrier concentration and chemical composition can be used to tune these materials through ferroelectric quantum transitions as shown in this figure adapted from 
Suski et al. \cite{Suski83}.{\bf c)} Quantum phase transition in a compositionally tuned organic uniaxial ferroelectric tris-sarcosine calcium chloride.  Here the quantum ferroelectric transition is tuned by chemical substitution.  This figure is adapted from Rowley et al. \cite{Rowley15}.  {\bf d)} Pressure-temperature phase diagrams of the charge-transfer complexes $DMTTF-QCl_4$ and $DMTTF-QBr_4$.  Inset:  Close to $P_c$, $T_c^2$ scales with $P$ in the ionic antiferroelectric $DMTTF-QBr_4$. We note that this scaling is similar to that of $T_c (x)$ shown in the inset of Figure 9, suggesting that external and chemical pressure have similar effects on $T_c$.  This figure is adapted from Horiuchi et al. \cite{Horiuchi03} with permission. 
}

%% file: Sec5_LTAs.tex
%7/10/17
%
%A Flavor for Low Temperature Applications

Let us now turn to some low-temperature applications of ferroelectrics.
As we mentioned earlier, the current trends due to market demands are for 
faster and smaller devices.  Ferroelectric films are used as passive
elements in dynamical random access memories (DRAMs) comprised 
of grids of capacitors with access transistors; 
here each bit is stored in a distinct capacitor where $0$ and $1$ 
correspond to the absence/presence of charge \cite{Scott00,Scott07} and 
the appeal of FE (and PE) materials is their high dielectric constants. 
DRAMS are among the highest density memories 
in current use with readily available 64 Gbit chips. Despite their 
many attractive features that include ultrafast speeds and low cost, 
DMRAMs require regular memory refresh cycles to ensure that the 
stored data is not lost due to everpresent leakage currents.  The 
refresh interval, currently about 60 milliseconds, depends
on the ratio of the stored charge to the leakage current.  An area
of current interest is to lengthen the time between refresh cycles,
both to increase device time for memory access and to reduce power 
consumption. If such a ``long-refresh DRAM''  were run
at 77 K, where the leakage currents are significantly smaller than
at ambient temperature, the refresh frequency might drop orders
of magnitude from kHz to Hz where details would depend on material
specifics. 

Ferroelectric films are also used as active memory elements in FeRAMs
(ferroelectric random access memories, also called FRAMS) 
where information is stored in polarization (charge) 
states \cite{Lines77,Scott00,Han13}. The low cost and high 
speed of FeRAMs makes them competitive with other storage 
devices \cite{Scott00,Han13} if they can maintain the demands
of miniaturization \cite{Scott98}; they are particularly
attractive for satellite applications due to their 
radiation hardness \cite{Scott00}.
Data storage cells in FeRAMs, as in DRAMs, consist of ferroelectric 
capacitor-based structures with access transistors; in FeRAMs it is
the nonlinear relationship between applied field and polarization (charge)
in ferroelectric materials that is exploited to store information 
analogous to the situation in
magnetic core memories.  For such a memory cell, the 
switching barrier ($\Delta U$) must be larger than the 
thermal energy scale, $k_B T$, so that the stored information is 
not corrupted.  Therefore we can equate the
switching and the thermal barriers
\begin{equation}
\Delta U =  k_B T   \quad\quad \Rightarrow \quad\quad L_c
\end{equation}  
to obtain a critical length-scale $L_c$ that sets the lower-bound
on the characteristic system size.
In a ferroelectric memory, the switching 
barrier can be estimated as the energy stored in its 
effective capacitor.  
Since these devices are operated at fixed 
voltage ($V$, the standard silicon logical level that is currently
$4.5 \pm 0.5$ volts) with effective capacitance $C$, we write
\begin{equation}
\Delta U = \frac{1}{2} C V^2 \propto L \quad\quad {\rm (for \quad fixed} 
\quad V)
\end{equation}
so that we see that the switching barrier scales with $C$
and hence with its characteristic length \cite{Scott00}.
More specifically, taking $C = (\epsilon_0 \epsilon) (\alpha )$
where $\alpha = \frac{A}{d}$, we find that
\begin{equation}
L_c = \left(\frac{T}{V^2}\right) \left(\frac{4}{\alpha \epsilon}\right) 10^{-13} \quad {\rm m}
\end{equation} 
where $T (K) $, $V(v) $, $\alpha (m)$ and $\epsilon$ are inputs.
A typical FRAM currently available uses PZT (lead zirconate titanate, 
$Pb(Zr,Ti)O_3$, with $\epsilon = 1300$) and
operates at ambient temperature 
($T$ = 300 K) with $\alpha = 10^{-5}$ m 
since it is 100 nm ($L$) thick with a lateral length of about 1 micron; at 
the current voltage standard (5 volts) , 
$L_c$ is 0.1 nm ($L \gg L_c$) indicating that these FRAMs 
are thermally safe.  However 
should $V$, $\alpha$ and/or $\epsilon$ decrease in the future, $T$ is a 
very useful tuning parameter that can be reduced to ensure 
that the stored charge is robust to thermal fluctuations.  In Figure \ref{fig11}
we show the scaling of the characteristic length $L_c$ for three specific
materials at room temperatures using current device parameters.

\fight=4in
\fg{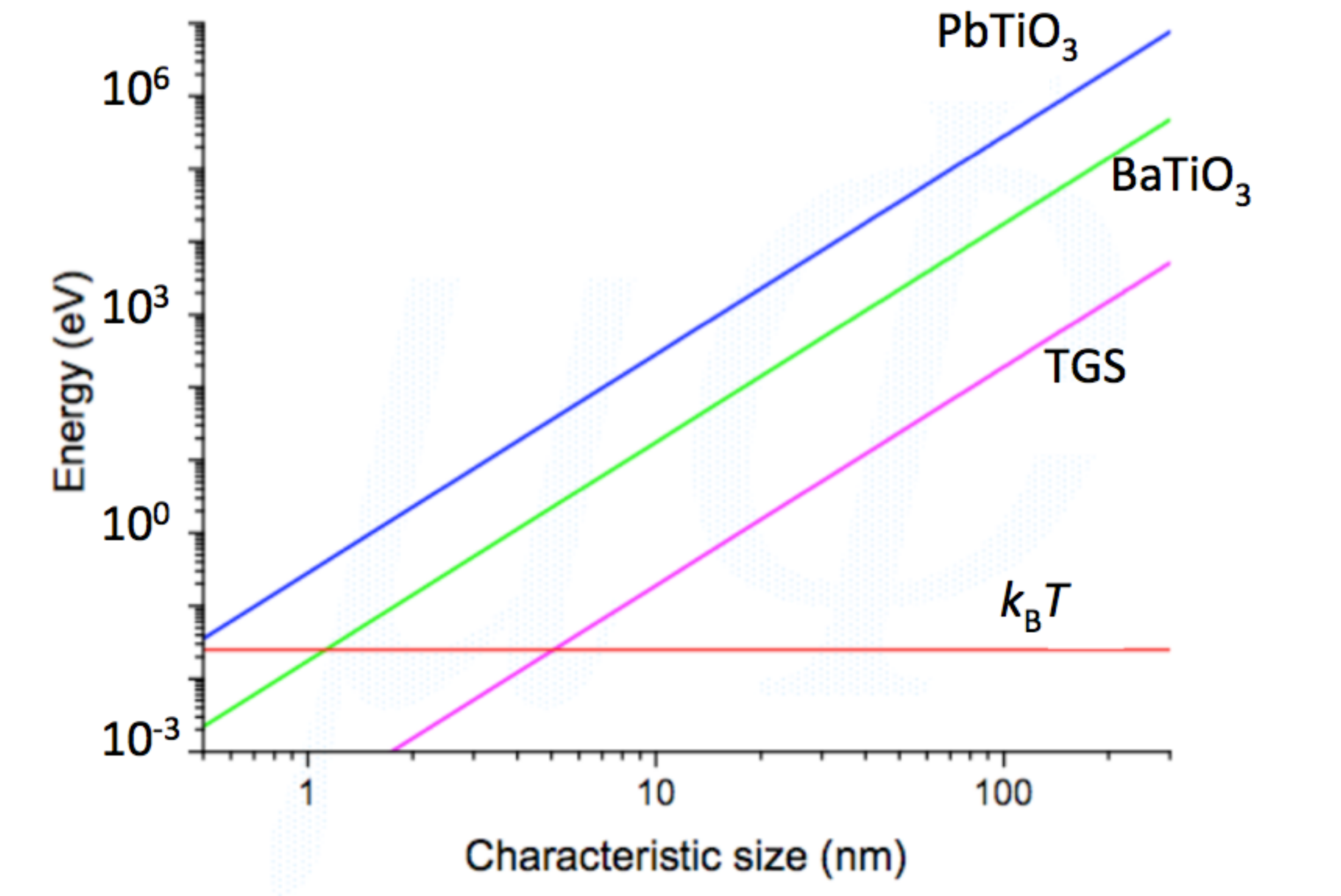}{fig11}{The minimum  device size for 
room-temperature operation without thermal corruption
for three different materials where TGS stands for triglycine sulfate; 
reproduced from M. Alexe \cite{Alexe03} with permission and thanks.}

Reduced operating temperatures leads to decreased conductivities 
and thus to increased breakdown fields \cite{Scott00}.  Higher $E$
fields can then be applied, resulting in increased charge and hence 
enhanced signal to noise for the sense amplifiers \cite{Scott00}; we 
recall that the relative polarization is the switched charge per 
unit area.  Typically this is determined by applying a series of 
voltage pulses before and after the switching.  The resulting 
currents are measured over time and and these integrated curves determine
the switched charge \cite{Scott00,Rabe07}. 
Because the voltage is fixed at a standard logic level, increased  
electric fields require decreasing the FE film thicknesses.
However if we try to increase stored charge by making a FE film 
thinner at room temperature, it may short since its conductivity is too high
to prevent breakdown.  More generally, the breakdown threshhold depends on 
the product of the electric field and the conductivity ($\sigma$) or
rather on the ratio $\frac{V\sigma}{d}$ \cite{Scott00}. Therefore for
fixed $V$, we can reduce the film thickness $d$ if we also decrease
$\sigma$ which is achieved by lowering the ambient temperature.  

Luckily ferroelectrics themselves can play a role in refrigeration via 
electrocaloric cooling (EC), the reduction of
temperature of a FE material in response to the removal of
an electric field \cite{Lines77,Jona93,Scott00,Lu09,Scott11,Takeuchi15}. 
Its magnetic counterpart, magnetocaloric cooling (MC), 
if often used to access millikelvin temperatures. Until recently EC 
effects were too small for practical applications and thus were not pursued.  
However several developments \cite{Lu09,Scott11,Takeuchi15,Mischenko06},  
suggest that we should revisit this phenomenon, particularly at 
low temperatures.  More specifically the breakdown fields of FE films 
are significantly larger than those of their bulk counterparts
so that higher $E$ fields can be applied, and multicapacitor technology can
be used to increase their effective volumes \cite{Scott11}. 
But we are getting ahead of ourselves.  In the spirit of
being self-contained, let's remind ourselves of the key features
of adiabatic cooling so that we can understand why to date the
magnetic version has been more successful than its electric
counterpart at low temperatures (and why we believe this topic 
deserves to be revisited!).
 
\fight=5in
\fg{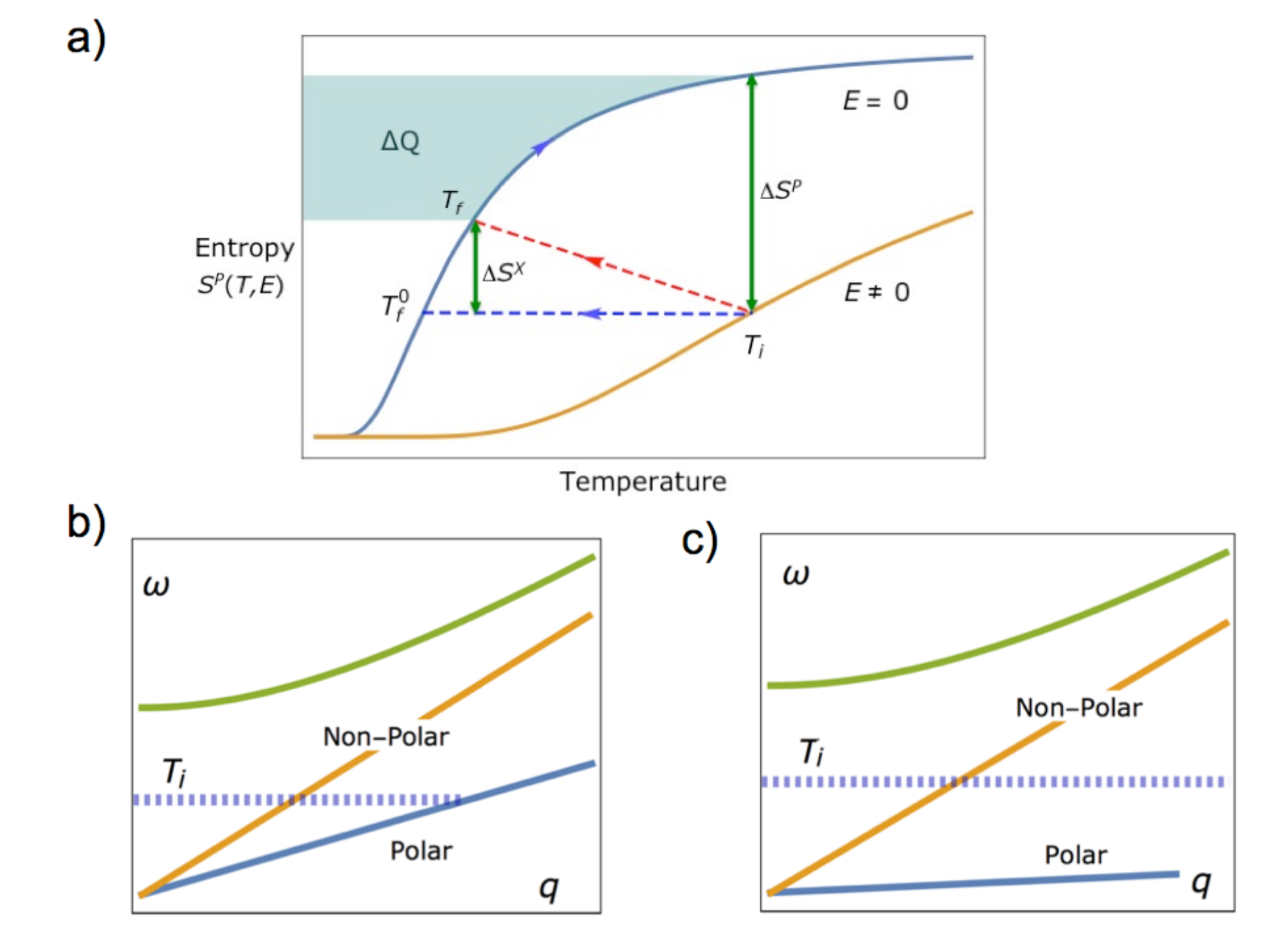}{fig12}{{\bf (a)} $S^P(T,E)$ Entropy-temperature
cycling for two distinct field strengths indicating the Carnot-like heat cycle
that is the basis for electrocaloric cooling. 
Here $T_i$ is the initial temperature in the adiabatic depolarization process,
$T_f^0$ is the final temperature in the absence of coupling to the non-polar
modes and $T_f$ is the final temperature including the effect of the 
non-polar modes. 
$\Delta Q$ is the heat that can be extracted from an external load.  
We require $\Delta S^P$ to be substantially greater than $\Delta S^X$ 
for effective cooling to occur (as in case {\bf (c)} in contrast to case 
{\bf (b)}.
{\bf (b)} Hypothetical dispersion where the sound velocity in the polar branch is less than that in the nonpolar branch. {\bf (c)} Hypothetical dispersion where the polar modes have a relatively flat dispersion, indicating very low interactions between the ions or the dipoles and thus large polar entropy.}

The entropy as a function of field and temperature ($S(E,T)$) plays
a key role in the electrocaloric effect and its magnetic analogue (MC) 
where $E$ is replaced by $B$. We can write
\begin{equation}
dS = \left( \frac{\partial S}{\partial T} \right)_E dT + 
\left( \frac{\partial S}{\partial E} \right)_T dE
\end{equation}
where, for an adiabatic process ($dS=0$) and using the Maxwell Relation
$\left( \frac{\partial S}{\partial E} \right)_T =  \left( \frac{\partial P}{\partial T} \right)_E $
(where $P$ is the polarization),
we obtain
\begin{equation}
- \left( \frac{\partial T}{\partial E} \right)_S 
= \frac{\left(\frac{\partial P}{\partial T} \right)_E}
{\left( \frac{\partial S}{\partial T} \right)_E} =  T \frac{\left(\frac{\partial P}{\partial T} \right)_E}{c_E}.
\label{ECEcoeff}
\end{equation}
Here $c_E$ is the specific heat at fixed electric field that has contributions from 
polar ($c_E^P$) and nonpolar ($c_E^X$) modes, where the latter are predominantly acoustic phonons.
The specific heat is of course a measure of the entropy and 
thus its magnitude will be related to the dispersion which together with the Bose function determines the
distribution of low-energy excitations as a function of wavevector
in the Brillouin zone.  In a displacive FE, the low-frequency polar modes are localized in q-space and $c_E^P$ 
is exponentially suppressed with a gap ($E \ne 0$); 
thus at low temperatures $c_E$ is dominated by $c_E^X$ 
and varies as $T^3$. These same acoustic phonons, in the absence of 
a ferroelectric phase transition, are the main contribution to the pyroelectric 
coefficient $\left(\frac{\partial P}{\partial T}\right)_E$;  
it is expected to decrease sufficiently rapidly with decreasing
temperature that $\left(\frac{\partial T}{\partial E}\right)_S$ in 
(\ref{ECEcoeff})
vanishes in the 
limit $T \rightarrow 0$ \cite{Lines77}.  Consistent with this
argument, cryogenic studies of $KTaO_3$ and $SrTiO_3$ yielded negligible 
EC effects \cite{Lawless77,Lawless77b,Radebaugh79,Radebaugh80}  
and this approach to low temperature refrigeration has not been actively pursued for some time.
So why then can magnetocaloric cooling be used routinely to access very low temperatures in 
complete contrast to its electric counterpart (to date)? 

We can address this question by looking at the entropy of
the polar modes, $S^P (E,T)$,  as a function of field and temperature
shown schematically in Figure \ref{fig12} a).  Here we start at an initial
temperature $T_i$ at $E=0$ and isothermally apply a finite
electric field; the entropy of the polar modes is then lowered.
The electric field is then removed adiabatically, and 
the temperature $T_f^P$ associated with the polar modes (uncoupled to other degrees of freedom) decreases. 
However since the total system is in equilibrium {\sl all} modes, polar
and nonpolar, must be at the same temperature. 
More specifically the total entropy ($S$) is a 
sum of the polar and the nonpolar contributions, $S = S^P + S^X$
and there will be overall cooling of the system if and only 
the entropy $\Delta S^P$ is substantially greater than $\Delta S^X$ in
Figure \ref{fig12} a:
%if the
%``take-home'' entropy
%\begin{equation}
%\Delta S = (S^P_f - S^P_i) - (S^X_f - S^X_i) = \Delta S^P - \Delta S^X > 0
%\end{equation}
%is finite and positive; 
%
%
%
more to the point, a nonzero 
$\Delta S^P$ is not good
enough! In other words, the polar entropy loss due to
the applied electric field must exceed the entropy to be removed from the
acoustic phonons;  in this case the system is cooled to
a final temperature $T_f$ such that $T_f^P < T_f < T_i$ 
as shown schematically in Figure \ref{fig12} a).   
This is difficult to achieve in simple 
displacive ferroelectrics where the sound velocity of the polar modes
is not substantially below that of the nonpolar ones.
One way to obtain $\Delta S^P >> \Delta S^X$ might be to reduce the sound velocity 
in the polar branch significantly compared
to its nonpolar counterpart (see Fig. \ref{fig12} b), effectively reducing
the coupling between electric dipoles to increase their entropy.
Another approach to $\Delta S^P >> \Delta S^X$ is to identify materials where 
the polar modes have flat dispersion bands, again indicating
low dipole-dipole effective interactions and a high polar entropy. 
(see Fig. \ref{fig12} c). We note that such flat dispersions are signature 
features of spin systems, specifically dilute paramagnetic salts and 
frustrated magnets, that are 
commonly used in cryogenic solid-state refrigeration \cite{Zhitomirsky03}. 
Because the dipole-dipole interaction is typically several orders 
of magnitude larger for electric dipoles than for their magnetic 
counterparts \cite{Chandra07}, the identification of paraelectric 
and ferroelectric materials with the necessary high polar 
entropy at low temperatures is particularly challenging.

What about electrocaloric cooling at low temperatures near
a ferroelectric quantum critical point?  Interestingly enough,
this question has already been posed near a magnetic quantum
critical point \cite{Wolf14}, and work is currently
in progress to study the FE case \cite{Chandra16}. Ideally
we want a system with a high density of minimally coupled
electric dipoles at low $T$ to achieve an enhanced polar entropy; possible
candidates include order-disorder, relaxor materials and ferrielectric
materials. Ideally we'd be approaching a quantum
tricritical point to maximize the change in polarization
without hysteresis; if we want the sound velocity of the polar
modes to approach zero, then we also want to be at a Lifshitz
point.  Furthermore we'd like the system to have a uniaxial polarization
to maximize coupling to the electric field ($\vec{E} \cdot \vec{P}$). 
Amnonia sulphate is an order-disorder ferroelectric with
a high entropy at its FE transition, though
it has not been practical for EC at room temperature due to its
ionic conductivity \cite{Scott11}.  This may not be an 
issue at low temperatures where ionic motion becomes 
frozen \cite{Scott11}. Indeed, analogous 
to their magnetic counterparts, dilute paraelectric salts
have been used to cool small samples to millikelvin 
temperatures \cite{Shepherd65,Shore66,Lawless69,Pohl69}; 
with current multicapacitor technology this technique could be greatly
improved and should be revisited.
 
In principle low-temperature electrocaloric cooling has many advantages over its
magnetic counterpart, particularly its reduced size (no magnets
necessary!) and its comparative simplicity of operation....we just
have to find the right materials to make it work!   Joule 
heating should not be a problem since the polar materials are reasonable
insulators. For space applications, where dilution refrigeration is
difficult to use particularly in microgravity conditions \cite{Wolf14,Shirron07}, 
electrocaloric cooling has an additional advantage as FE materials are 
robust to everpresent cosmic rays \cite{Scott00}.

Other possible applications for low-temperature paraelectric/ferroelectric
materials include:

\begin{itemize}

\item{\bf Satellite Electronics.}  The radiation effects, due to cosmic
rays and to solar activity, are not evenly distributed for low-Earth
orbits and are even harsher in outer space.  There is an urgent need for
new electronics that are high-performance, radiation-tolerant and 
reliable \cite{Sayyah11} at an ambient temperature of roughly $10K$,
and onboard infrared detectors require $mK$ operating temperatures.

\item{\bf Phased-Array Radar.}
Ferroelectric-superconductor ``sandwiches'' hold promise as phase shifters
in phased-array radar GHz devices, running at significantly lower voltages than
current versions.  The dielectric losses must be kept very low to
be competitive with existing bulky technologies and thus they would have to
be run at low temperatures \cite{Scott00}, possibly maintained by 
electrocaloric cooling.

\item{\bf High Permittivity Supercapacitors.} There is an increasing
need for high density storage of electrochemical energy with rapid 
charge/discharge cycles and long lifetimes.  Low dielectric loss
and large-scale requirements could make this a niche for low-temperature 
PE/FE materials that are relatively cost effective \cite{Hoffmann15}

\end{itemize}

and there are certainly many more!

%% file: Sec6_OpenQs.tex
%7/10/17
%
%Open Questions for Future Research

In order to emphasize research prospects, we conclude with a list of 
open research questions in this area of materials near ferroelectric quantum phase transitions:

\vskip0.2in

\begin{itemize}

\item {\bf Specific FE Materials for Study at Low Temperatures.}
\vskip0.10in
Here we have argued  that the study of materials near their 
ferroelectric quantum critical points (FE-QCPs) can play
an important role towards understanding universality at 
quantum phase transitions. However there are only a few systems 
currently known that remain paraelectric to the lowest temperatures, 
so are those the only materials in this class to study?   
There are certainly many materials with low (classical) ferroelectric 
transition temperatures ($T_c < 100 K$) \cite{Lines77,Scott00}, and  
we expect that these $T_c$'s could be reduced with pressure, stress 
or with chemical or isotopic substitution to yield possible QCPs that, 
to our knowledge, have not yet been explored. Empirically it seems that
ferroelectric transition temperatures are very sensitive to pressure, as
shown in Figure (\ref{figxx} a) with the case of $BaTiO_3$.  If this 
pressure-sensitivity of $T_c$ is indeed the general case, then this would 
significantly broaden the range of materials \cite{Lines77,Scott00} where 
quantum phase transitions could be studied.  Furthermore the possibility
of antiferroelectric quantum criticality could be pursued in materials
like $NaNbO_3$ with coexisting ferroelectric and 
antiferroelectric interactions \cite{Mishra07} whose low antiferroelectric $T_c$ ($\sim 12 K$) 
could be reduced (e.g. by substitution) and where quantum fluctuations 
are known to be important at low temperatures \cite{Raevskaya08}.
We note that competing energy- and length-scales can lead to quantum
electric-dipole liquids \cite{Shen16}, novel textures \cite{Scott05,Naumov04,Dawber06,Goncalves-Ferreira08} and exotic topological excitations \cite{Rowley14b,Lin14} in the vicinity of these quantum phase transitions.  

\vskip0.3in

%\input boxes/LowT_FEs

%\vskip0.750in

\item{\bf Add Spin:  A Multiferroic QCP.} 
\vskip0.10in
Additional degrees of freedom can be added in a systematic fashion
to materials near their ferroelectric quantum phase transitions with rich
phase behavior expected \cite{Scott15}.   
For example, quantum criticality in  
multiferroic materials \cite{Dong15} is only 
starting to be explored 
\cite{Morice16,Katsufuji01,Kim09,Das12,Schiemer16,Rowley16b} 
where the possible 
interaction of two quantum critical points could lead to novel behavior. 
Of course here we have been predominantly discussing bulk materials, but
the low temperature behavior of multiferroic heterostructures \cite{Vaz10,Vaz12} could
be intriguing as well.  
Multiferroics 
at low temperatures with high polar and spin entropies could also be 
candidates for advanced cryogenic solid-state refrigeration \cite{Midya16}
based on both the 
electrocaloric and the magnetocaloric effects.  We also note the intriguing case of multiferroic relaxor quantum critical points 
\cite{Rowley15,Rowley16b}, that may be related to quantum glassiness.

\vskip0.30in

\item{\bf Add Charge:  An Exotic Metal and Unusual Superconductivity}
\vskip0.10in
The study of quantum criticality in magnetic metals is often motivated by
the search for non-Fermi liquids and for unconventional 
superconductivity \cite{Gegenwart08}.
It is thus fitting that we note that the study of materials near a FE-QCP also
fits into this ``grand scheme.''

\vskip0.15in

Charge is another degree of freedom that can be added to a material near its 
FE-QCP by either chemical and/or gate doping.  The Mott 
criterion \cite{Mott90} for the 
critical dopant concentration ($n_c$) for a 
metal-insulator transition in doped semiconductors occurs when the 
average dopant-dopant distance ($d = n^{-\frac{1}{3}}$) 
is a significant fraction of the effective
Bohr radius ($a^*_B = \frac{\epsilon \hbar^2}{m^* e^2}$) 
where $\epsilon$ is the 
dielectric constant; 
more concretely the critical 
concentration $n_c$ is defined as $n_c^{\frac{1}{3}} a_B^* \approx 0.26$, 
consistent with experiment in many 
semiconductors \cite{Edwards78}.  Since the effective Bohr radius is proportional to the 
dielectric constant ($\epsilon$), it is much larger in n-doped STO than in doped
semiconductors based on silicon or germanium (see Figure \ref{fig13}); therefore a lower
$n_c$ is expected, consistent with observation \cite{Lin13,Edwards78}.

\fight=4in
\fg{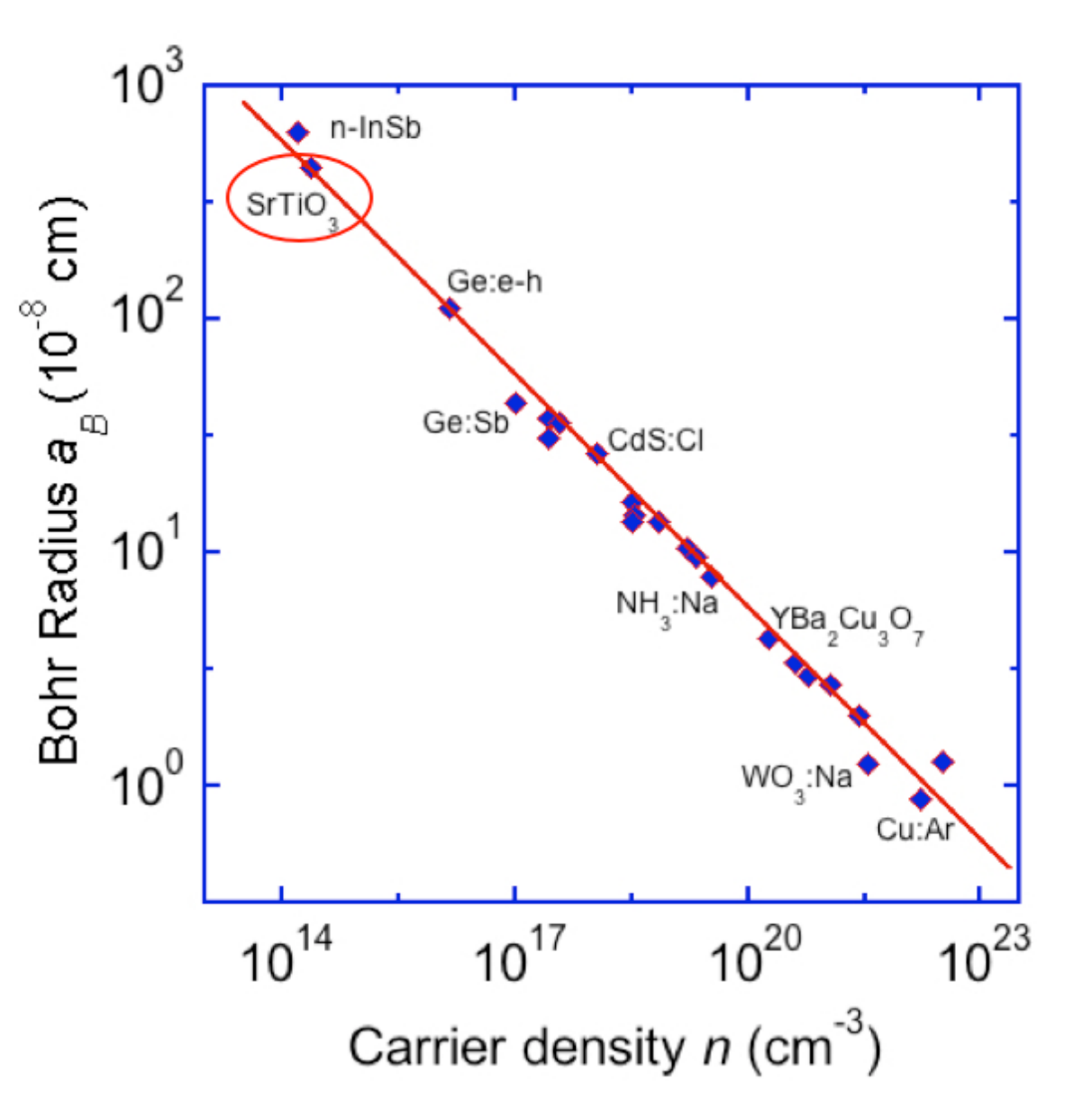}{fig13}{A plot of the effective Bohr radius ($a_B$) vs. 
carrier density ($n$) indicating good comparison between the Mott criterion ($n_c^{\frac{1}{3}} a_B = 0.26 $)
for the metal-insulator transition 
and experimental systems where $a_B$ and the critical carrier density ($n_c$) for metallicity are known.
Because the effective Bohr radius is inversely proportional to the dielectric constant, it is large
for $SrTiO_3$ indicating a low critical carrier concentration for the metal-insulator transition consistent
with observation \cite{Lin13}.  This figure is adapted from 
Edwards and Sienko \cite{Edwards78} with permission 
and with thanks to K. Behnia \cite{Behnia16}.}

\vskip0.15in

The Fermi temperature of metallic n-doped STO can be quite low because of the
relatively high carrier effective mass and low densities of practical
interest; for example for 
$n = 5.5 \times 10^{17} cm^{-3}$, $T_F \approx 13 K$ \cite{Lin13}.
At first sight this dilute-carrier metal looks quite conventional with a 
resistivity that scales like $T^2$ as expected for a three-dimensional 
Fermi liquid \cite{Okuda01}.  The catch is that this behavior continues 
to temperatures 
well above the Fermi temperature $T_F$ \cite{Lin15,Okuda01}
where $T_F$ is determined from the coefficient of the linear heat 
capacity; in 3d for fixed $m$, $T_F$ scales with $n^{\frac{2}{3}}$.  
Arguments based on Fermi liquid, requiring that $T << T_F$, are
clearly inapplicable for $T > T_F$; furthermore $A$, 
the coefficient of this $T^2$ behavior in the resistivity, 
can change by four orders of 
magnitude by tuning the carrier concentration and persists to
dilute limits where known mechanisms for $T^2$ behavior are no longer 
applicable \cite{Lin15}.  

\vskip0.15in

The traditional BCS theory of superconductivity \cite{Schrieffer83} requires 
$T_F \gg T_D$ where $T_F$ and $T_D$ are the Fermi and the Debye temperatures, 
a condition not satisfied in n-doped STO; for 
$n = 5.5 \times 10^{17} cm^{-3}$, $T_F \approx 13 K$ and $T_D \sim 400 K$ 
so that $T_F \ll T_D$ \cite{Lin13}.  The possibility of superconductivity in 
doped paraelectric  materials was considered within the decade
after the BCS theory was developed \cite{Gurevich62}, and it was originally 
suggested that in polar semiconductors the temperature-scale 
associated with the longitudinal optical phonon, $T_L$, could 
replace $T_D$ in the BCS formalism. However, because typically
$T_F \ll T_L$ for densities $n \lesssim 10^{19} cm^{-3}$, the 
implication was that superconductivity in doped paraelectrics 
was unlikely \cite{Gurevich62}.  
Nevertheless superconductivity was 
predicted \cite{Cohen64a,Cohen64b} in n-doped STO based on
intervalley scattering; this
theory led to the experimental search and subsequent
observation of superconductivity \cite{Schooley64} in this material.
Ironically, despite this finding, it was later shown that key aspects of
the motivating theory, particularly the assumption of multiple valleys,
were inapplicable to STO \cite{Cohen69}; this unusual 
twist in the discovery
of superconductivity in doped STO only makes its existence 
all the more remarkable \cite{Edge15,Kedem16,Fernandes13,Ruhman16,Gorkov16a,Gorkov16b}.     

\vskip0.15in

%\fight=5in
%\fg{figs/Fig14.pdf}{fig14}{Superconductivity on the Border of
%Ferroelectricity in bulk Electron-Doped STO (Adapted from Kooce NEED TO INCLUDE REFERENCE HERE BUT DOUBLE-CHECK WITH GIL (with permsssion)).}

In summary superconductivity occurs in n-doped STO, and we 
still have a lot to learn about its underlying mechanism and the symmetry of 
its order parameter.
It has been observed both in bulk \cite{Rischau17,Lin13,Stucky16,Schooley65,Koonce67,Binnig80,Suzuki96}
and, more recently, 
at the interface of $LaAlAs/SrTiO_3$ \cite{Biscaras10,Moetakef12,Pai17}.
Like many of the heavy fermion superconductors, it is in the parameter regime $T_F << T_D$ and thus 
cannot be described by conventional BCS theory; however here spin-fluctuation mediated pairing 
cannot be applied.  Instead it is natural to consider electron-electron interactions mediated by
long-range Coulomb potentials.  
However here there is a conundrum:  the pairing interaction $V (\omega)$ 
scales inversely proportional to the dielectric constant $\epsilon (\omega)$ so that at $\omega = 0$ the interaction
is small (since $\epsilon (0)$ is large).  We recall that, within a soft mode picture described by (\ref{osc}) and (\ref{LST}), the dielectric constant can be
written as 
\begin{equation}
\frac{\epsilon (\omega)}{\epsilon_\infty} = 1 - 
(\omega^2_{LO} - \omega^2_{TO}) \frac{\omega_{TO}^2}{(\omega^2 - \omega_{TO}^2)}
\end{equation}      
where the transverse and longitudinal frequencies, 
$\omega_{TO}$ and $\omega_{LO}$, are defined by the zero and the 
pole of $\epsilon (\omega)$.  
We see that in the frequency window
\begin{equation}
\omega_{TO} \ < \ \omega \ < \ \omega_{LO}
\end{equation}
$\epsilon (\omega)$ is negative leading to an {\sl attractive} interaction $V(\omega)$; furthermore we note that 
this ``attractive frequency range'' is increased to its maximal value close to a 
FE-QCP where $\omega_T \rightarrow 0$.  Here, for simplicity, we have suppressed the $q$-dependence of $V(\omega)$ 
and $\epsilon(\omega)$,
but it is likely to be important due to the long-range nature of the Coulomb interaction.  Furthermore we need to consider screening effects of the added
carriers that become progressively more important with increasing $n$.

\vskip0.15in

So here we have a dynamical interaction between the electrons...what's so difficult about
this superconducting problem?  Actually there are two challenges to address.  The first
is that a key aspect of Cooper's crucial superconducting pairing argument relies on being close to the Fermi
energy \cite{Schrieffer83}; in this case the pairing problem becomes effectively 2d where, in contrast to 3d,
binding is possible with an arbitrarily weak attraction.  This reasoning is not applicable to n-doped STO
where the pairing energy-scale is much higher than $T_F$.  Second, any attractive pairing of electrons must somehow ``bypass''
their repulsive Coulomb interaction. In the BCS theory retardation is 
crucial \cite{Schrieffer83}:  
the ionic screening cloud lags behind the electron, thereby mediating its attraction to other electrons. 
By contrast in n-doped STO, where there is no similar large separation of time-scales, further study of possible 
``Coulomb circumvention'' mechanisms is needed. In a nutshell in superconducting n-doped STO we are without 
two key features of the successful
BCS theory of superconductivity...how would the theoretical description of 
superconductivity have developed if this amazing phenomenon
had first been observed in n-doped STO rather than in mercury?

\vskip0.15in

%\fight=6in
%\fg{figs/Fig15.pdf}{fig15}{Four materials that, for different reasons,
%have large effective Bohr radii and high mobilities; they all, when doped
%go superconducting !(adapted after K. Behnia (talk?)
%with permission and thanks).}

Finally we should note that electron-doped STO is one of the most dilute 
superconductor known to date \cite{Rischau17,Lin13}; its 
density of charge carriers, coming from 
niobium doping (on Ti sites), lanthanum substitution (on Sr sites) or from oxygen vacancies, is comparable to that of 
the metal bismuth that has only very recently been
shown to go superconducting, albeit at a temperature much lower than that
observed in n-doped STO \cite{Prakash16}. 
%Interestingly enough we know of four materials, including n-doped STO,
%that for different reasons have large effective Bohr radii and high mobilities (see Figure 15) ...and 
%all four superconduct!

\vskip0.15in

Since much study of quantum criticality is motivated by the search for novel forms of superconductivity, let us note another research possibility 
in this direction.  Doped strained STO is a good candidate for a polar metal and indeed is currently a topic of active study in multi-component 
metallic/dielectric heterostructures where STO is known to host a finite polarization \cite{Park17}.  Though such polar metals 
were predicted theoretically some time ago \cite{Anderson65}, recently there has been a resurgence of interest in such materials in 
part due to their anisotropic thermal and magnetoelectric properties \cite{Benedek16,Lezaic16,Kim16}.  At low temperatures such polar metals will 
surely become polar superconductors; such non-centrosymmetric superconductors are expected to
have mixed-parity pairing mechanisms with topological aspects to their superconducting states \cite{Yip14}.

\end{itemize}

\vskip0.20in

These are just some of the many research questions that emerge from 
looking at paraelectrics and ferroelectrics at low temperatures; proximity to 
quantum phase transitions can be tuned by either pressure, stress,  
chemical or isotope substitution and perhaps even more.  This is a 
rich area with plenty to explore, and we look forward to progress
in these and many related topics.  
%We look 
%forward to writing (or reading) a review in the near future 
%that includes progress in these and many related topics!

%% file: Sec7_Acks.tex
We have benefitted from discussions with many colleagues including M. Alexe, 
K. Behnia, T. Birol, P. Coleman, M. Continentino, D. Khmelnitskii, 
P.B. Littlewood, L. Palova, K.M. Rabe, S.S. Saxena, Q. Si and C. Yu 
as this article has evolved. 
PC acknowledges Trinity College, Cambridge  and the Cavendish
Laboratory where this project was initiated. This work was supported  
by National Science Foundation grant NSF-DMR-1334428 (PC). 
GGL acknowledges support from grant no. EP/K012894/1 of the EPSRC and the 
CNPq/Science without Borders Program, and SER acknowledges support from a CONFAP Newton grant.
PC and GGL also thank the Aspen Center for Physics and the 
National Science Foundation Grant No. PHYS-1066293 for hospitality where 
this work was further developed and discussed.

%\vskip02.0in